\documentclass[journal]{IEEEtran}
\usepackage[swedish,english]{babel}
\usepackage{microtype}

\RequirePackage{etex} 
\usepackage{graphicx,amsmath,url,amssymb,mathtools,steinmetz,bm}
\usepackage[hidelinks]{hyperref} 
\usepackage[nameinlink,capitalise]{cleveref}
\crefname{thm}{Theorem}{Theorems}
\crefname{equation}{}{}
\usepackage{bbold}
\usepackage[caption=false,font=footnotesize]{subfig}
\usepackage{cancel}
\usepackage{siunitx}
\DeclareSIUnit\perunit{p.u.}
\DeclareSIUnit\voltampere{VA}
\DeclareSIUnit\wattsecond{Ws}
\DeclareSIUnit\Newtonmeter{Nm}
\DeclareSIUnit\kgmeter{kg{\ensuremath{\,}}m}
\DeclareSIUnit\rpm{rpm}

\usepackage{accents}

\usepackage[usenames, dvipsnames]{color}

\newtheorem{rem}{Remark}

\newcommand{\complex}{\ensuremath{\mathbb{C}}}

\newcommand{\jimag}{\ensuremath{j}}
\newcommand{\jomega}{\ensuremath{\jimag \omega}}

\newcommand{\rated}{\ensuremath{\textrm{nom}}}

\newcommand{\optimal}{\ensuremath{\textrm{opt}}}
\newcommand{\MPP}{\ensuremath{\textrm{MPP}}}
\newcommand{\wind}{\ensuremath{\textrm{wind}}}

\newcommand{\FCR}{\ensuremath{ \textrm{FCR}}}

\usepackage{tikz}

\usepackage{matlab-prettifier}
\usepackage{scalerel}
\usepackage{tikz}
\usetikzlibrary{svg.path}

\definecolor{orcidlogocol}{HTML}{A6CE39}
\tikzset{
	orcidlogo/.pic={
		\fill[orcidlogocol] svg{M256,128c0,70.7-57.3,128-128,128C57.3,256,0,198.7,0,128C0,57.3,57.3,0,128,0C198.7,0,256,57.3,256,128z};
		\fill[white] svg{M86.3,186.2H70.9V79.1h15.4v48.4V186.2z}
		svg{M108.9,79.1h41.6c39.6,0,57,28.3,57,53.6c0,27.5-21.5,53.6-56.8,53.6h-41.8V79.1z M124.3,172.4h24.5c34.9,0,42.9-26.5,42.9-39.7c0-21.5-13.7-39.7-43.7-39.7h-23.7V172.4z}
		svg{M88.7,56.8c0,5.5-4.5,10.1-10.1,10.1c-5.6,0-10.1-4.6-10.1-10.1c0-5.6,4.5-10.1,10.1-10.1C84.2,46.7,88.7,51.3,88.7,56.8z};
	}
}

\newcommand\orcidicon[1]{\href{https://orcid.org/#1}{\mbox{\scalerel*{
				\begin{tikzpicture}[yscale=-1,transform shape]
				\pic{orcidlogo};
				\end{tikzpicture}
			}{|}}}}

\usepackage{cite}

\usepackage{soul}

\begin{document}

\frenchspacing

\title{
Variable-Speed  Wind Turbine Control \\Designed 
for Coordinated Fast Frequency Reserves 
}

\author{
	Joakim~Bj\"ork${\textsuperscript{\orcidicon{0000-0003-0656-7991}}}$,~\IEEEmembership{Member,~IEEE,}
	Daniel~V\'azquez~Pombo${\textsuperscript{\orcidicon{0000-0001-5664-9421}}}$,~\IEEEmembership{Student~Member,~IEEE,}
	and~Karl~Henrik~Johansson${\textsuperscript{\orcidicon{0000-0001-9940-5929}}}$,~\IEEEmembership{Fellow,~IEEE}%<-this % stops a space

%\thanks{This work has been submitted to the IEEE for possible publication. Copyright may be transferred without notice, after which this version may no longer be accessible}%
\thanks{\textcopyright~2021 IEEE.  Personal use of this material is permitted.  Permission from IEEE must be obtained for all other uses, in any current or future media, including reprinting/republishing this material for advertising or promotional purposes, creating new collective works, for resale or redistribution to servers or lists, or reuse of any copyrighted component of this work in other works.}%
\thanks{This work was supported by the KTH PhD program in the digitalization of electric power engineering and in part by the Knut and Alice Wallenberg Foundation, the Swedish Research Council, and the Swedish Foundation for Strategic Research.
}%
	\thanks{J. Bj\"ork and K. H. Johansson are with the School of Electrical Engineering and Computer Science, KTH Royal Institute of Technology, \mbox{100 44 Stockholm,} Sweden (email: joakbj@kth.se; kallej@kth.se).}% <-this % stops
	\thanks{D. V. Pombo is with the Department of Electrical Engineering at the Technical University of Denmark (DTU), Roskilde, Denmark and with Research and Development in Vattenfall AB in Stockholm, Sweden (email: dvapo@elektro.dtu.dk).}}

\maketitle

\begin{abstract}
Modern power systems present low levels of inertia due to the growing shares of converter-interfaced generation. Consequently, renewable energy sources are increasingly requested to provide frequency support. In addition, due to the inertia loss, the requirements regarding frequency containment reserves (FCR) are becoming tough to meet with traditional units such as hydro, whose non-minimum phase (NMP) characteristic reduces the closed-loop stability margins. The shortcomings of traditional synchronous generation motivates new protocols for fast frequency reserves (FFR). In this work, we design a wind turbine (WT) model useful for FFR. It is shown that the dynamical shortcomings of the WT, in providing steady-power or slow FCR support, are suitably described by a first-order transfer function with a slow NMP zero. The WT model is tested in a 5-machine representation of the Nordic synchronous grid. It is shown that the NMP model is useful for designing a controller that coordinates FFR from wind with slow FCR from hydro turbines. By simulating the disconnection of a \SI{1400}{\mega\watt} importing dc link in a detailed nonlinear model, it is shown that the wind--hydro combination not only satisfies the latest regulations, but also presents a smooth response avoiding overshoot and secondary frequency dips during frequency recovery.
\end{abstract}


\IEEEpeerreviewmaketitle

%%%%%%%%%%%%%%%%%%%%%%%%%%%%%%%%%%%%%%%%%%%%%%%%%%%%%%%%%%%%

\section{Introduction}
\IEEEPARstart{P}{ower} systems exhibiting low rotational inertia present faster frequency dynamics, making frequency control and  system operation more challenging.
Unlike conventional synchronous generation, converter-interfaced generation such as wind or solar does not contribute to the inertia of the grid.
As renewable production begins to replace conventional production, frequency stability is a growing challenge for the modern grid~\cite{milanoFoundationsChallengesLowInertia2018}.
A number of relatively recent blackouts are related to large frequency disturbances. The incidence of this phenomenon is expected to increase in the future as the energy transition continues; in fact they have doubled from the early 2000s~\cite{rahmanLargestBlackoutsWorld2016}.
% A number of cases have been studied in recent literature which can in general by grouped attending to the root cause as:
Examples from the literature attribute the root causes of recent blackouts to:
overloading of transmission lines following an unsuccessful clearing of a short circuit fault~\cite{ucteinvestigationcommitteeFinalReportInvestigation2004}, damage to transmission lines due to extreme weather~\cite{aemoBlackSystemSAustralia2017}, power plant tripping due to malfunctioning of protections~\cite{beckGlobalBlackoutsLessons2005}. In all of them,  the lack of frequency response from converter-interfaced renewable production made the system operators (SOs) incapable of avoiding blackouts.
With growing shares of renewables, SOs are therefore increasingly demanding renewable generation to participate in frequency containment reserves (FCR)~\cite{brundlingerReviewAssessmentLatest2016}.

Aggregating groups of small-scale producers and consumers together as virtual power plants is a proposed solution to allow smaller players with more variable production to participate in the market with the functionality of a traditional power plant~\cite{sabooriVirtualPowerPlant2011,ghavidelReviewVirtualPower2016}. The main objective is often to coordinate dispatch, maximize the revenue, and to reduce the financial risk of stochastic generation, such as wind, in the day-ahead and intraday markets~\cite{vasiraniAgentBasedApproachVirtual2013,wangInteractiveDispatchModes2016,alvarezGenericStorageModel2019}. But also other services, such as voltage regulation~\cite{moutisVoltageRegulationSupport2018} and allocation of FCR resources~\cite{zhongImpactVirtualPower2020,alhelouPrimaryFrequencyResponse2020} have been proposed.
In this work, we will design controllers for dynamic coordination of FCR over all frequency ranges/time scales, using the design procedure developed in~\cite{bjorkDynamicVirtualPowerunpublished,bjorkFundamentalControlPerformance2021}, to form a dynamic virtual power plant (DVPP). The purpose of the DVPP is to coordinate individual devices at the transmission grid level so that their aggregated output fulfills the SO's requirements, while accounting for the individual constraints of participating devices~\cite{posytyfConceptObjectives2021_manual}.

In this work, we consider fast frequency reserves (FFR) based on the definition used by the European Network of Transmission System Operators for Electricity (ENTSO-E). FFR differs from FCR in that it is not intended to stay on for a sustained period of time. The main objective of FFR is instead to provide a fast response to disturbances, acting as a complement to FCR for around 5--\SI{30}{\second}. It does not reduce the need of FCR, thus cannot replace FCR~\cite{entso-eFastFrequencyReserve2019}. FFR is therefore ideally provided by fast acting converter-interfaced generation, and may be implemented by sources that cannot sustain a prolonged power injection.

For wind turbines (WTs), the concept of synthetic inertia has gained a lot of attention. 
Briefly, synthetic inertia allows to temporally increase the output power of a WT in exchange for reducing the rotor speed. If the turbine is operated at its maximum power point (MPP), this necessitates a recovery period in which the output power is less than at the starting point, until the rotor speeds up again~\cite{morrenWindTurbinesEmulating2006,ochoaFastfrequencyResponseProvided2017,ullahTemporaryPrimaryFrequency2008,liCoordinatedControlStrategies2017,leeReleasableKineticEnergyBased2016,zhaoFastFrequencySupport2020} if not operated below the MPP~\cite{wilches-bernalFundamentalStudyApplying2016}. This means that MPP tracking WTs are unable to provide FCR for a sustained period of time. However, the speed at which the converter-interfaced WTs are able to react makes them suitable for providing FFR to complement conventional synchronous generation limited by the dynamic constraints of mechanical valves, servo systems, etc.~\cite{entso-eFastFrequencyReserve2019}. 
In~\cite{morrenWindTurbinesEmulating2006}, a variable-speed controlled WT is fitted with a proportional-derivative FFR controller. The controller activates when the frequency falls following the disconnection of a generator. By increasing the WT power output, the controller manages to reduce the system nadir. The FFR is deactivated after \SI{25}{\second} in order to restore the turbine to the MPP. 
In~\cite{ochoaFastfrequencyResponseProvided2017}, FFR is provided by shifting the MPP tracking curve based on frequency measurements.
The capability of a WT in  providing FFR is investigated in~\cite{ullahTemporaryPrimaryFrequency2008}. The proposed control method increases the output to a fixed level until the WT rotor speed falls below a low speed limit, after which the rotor speed must be restored to the MPP. In~\cite{liCoordinatedControlStrategies2017}, a controller for an offshore wind farm is designed to take advantage of the energy stored in the dc capacitors, reducing unnecessary deviations from the MPP when providing FFR. 
A challenge with providing FFR from wind is the second frequency dip that may follow when restoring the WT rotor speed. If the controller is not tuned correctly, this phenomenon may lead to an even worse disturbance response than FCR without wind-FFR and is therefore best avoided. 
% This is particularly important in low-inertia power systems. 
One way to circumvent this issue is to delay the rotor speed restoration 
~\cite{leeReleasableKineticEnergyBased2016,zhaoFastFrequencySupport2020}.
The drawback of such a method (as well as methods that use curtailment~\cite{wilches-bernalFundamentalStudyApplying2016}) is that the WT will spend more time below the MPP, resulting in lower energy production and economic loss. The production loss has to be covered by other sources. In this work, we will therefore consider MPP tracking WTs that immediately restore the rotor speed to the MPP. 

The available literature is mainly focused on what can be achieved with WTs in terms of synthetic inertia. It does not directly address the needs of low-inertia power systems.
As a result FFR may give unsatisfactory results, such as a second frequency dip when having to restore the WT rotor speed. In this work, we take a different approach. First, we specify the requirements of all participating actuators in terms of FCR and FFR based on the SO's requirements. Then, we identify the dynamic characteristics of the WTs (and other controlled actuators) that are needed to determine if the frequency control objective is feasible or not.

The contributions of this work are, firstly, the design of a variable-speed feedback controller, allowing a WT to provide FFR without curtailment. The design is similar to the control implemented in~\cite{morrenWindTurbinesEmulating2006}, but here, {the purpose is to enable a simplified model that describes the relevant dynamics for FFR}. When extracting power above the MPP, the turbine decelerates, thereby reducing its steady-output. Consequently, the FFR control action puts the WT in a unstable mode of operation. To amend this, a model-based stabilizing controller is implemented, using wind speed measurement and feedback of the normalized rotor speed. The controller is designed so that the WT exhibits similar dynamic properties for various wind speed conditions. Secondly, a simplified first-order linear approximation is derived.  The simplified  model is designed to be a ``worst-case" model, capturing the dynamic shortcomings relevant for safe coordination of FCR and FFR. This makes it possible to use tools from linear control theory to determine if the frequency control objective is feasible. Lastly, the simplified WT model is used for coordinated FCR and FFR design, using the model matching approach developed in~\cite{bjorkDynamicVirtualPowerunpublished,bjorkFundamentalControlPerformance2021}. The modified WT and the coordinated control are tested in detailed nonlinear power system simulations in Simulink Simscape Electrical~\cite{hydro-quebecSimscapeElectricalReference2020_manual}, both in a local DVPP and in a model of the Nordic synchronous grid. To study the control limitation for wind-FFR, it is more interesting to consider low-inertia power systems where requirements for FFR are more critical. The Nordic grid is an extra challenging case study since its FCR is almost exclusively provided by hydropower. Hydro has non-minimum phase (NMP) zero dynamics that gives it an initially inverse response when increasing its power output. This limits the achievable response speed and also further increase the need for FFR. These are the reasons why the Nordic synchronous grid is chosen for the simulation study. The results in this work are however applicable for general power systems.  

The remainder of the paper is structured as follows: \cref{sec:SOA} presents the problem formulation. In \cref{sec:WTmodel}, a benchmark \SI{5}{\mega\watt} WT model is adapted in order to capture relevant dynamics related to FFR. In \cref{sec:PowerSystem}, it is shown how the modified WT model can be used to coordinate wind power with other resources. Lastly, \cref{sec:Conclusion} concludes the paper with a discussion of the results.

\section{Background and Problem Formulation} \label{sec:SOA}
This section introduces the original National Renewable Energy Laboratory (NREL) \SI{5}{\mega\watt} WT model along with the considered Nordic 5-machine (N5) test system.

\subsection{The NREL \SI{5}{\mega\watt} Baseline WT Model} 
\label{sec:intro_NREL}
\begin{figure*}[tb!]
        \captionsetup[subfloat]{farskip=0pt}
		\centering
		\subfloat[\label{fig:simplified_NREL_model} Simplified block diagram.]
        {{\includegraphics[scale=0.75]{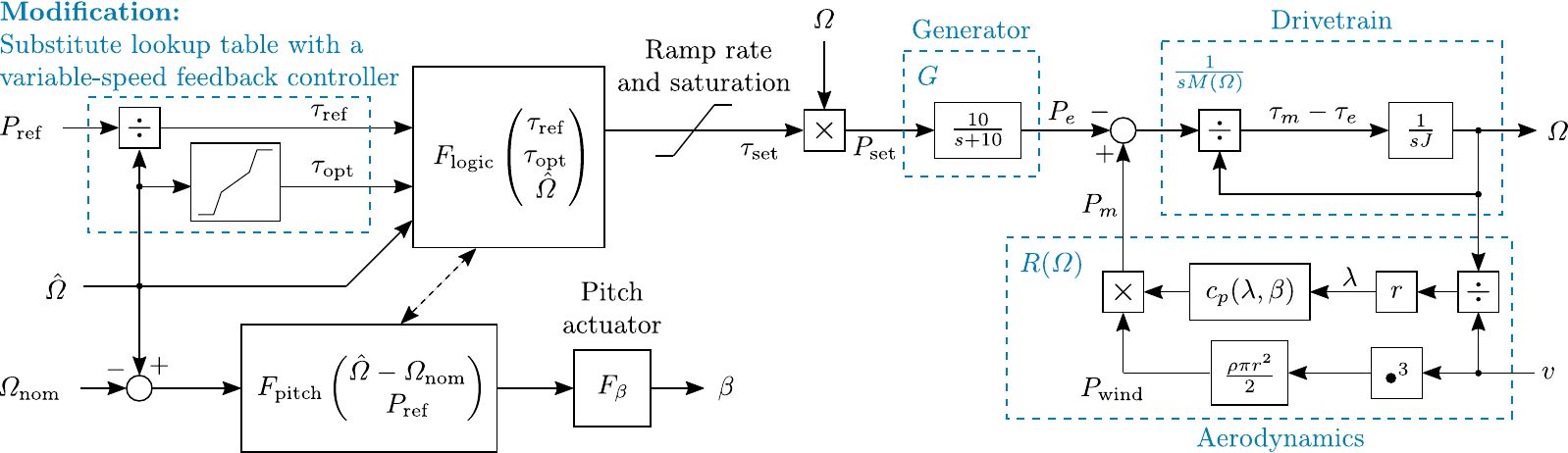}}}
		\hfill
		\subfloat[\label{fig:wind_speed_curves} Power/speed characteristic with $\beta=0$.]
        {{\includegraphics[scale = 0.7]{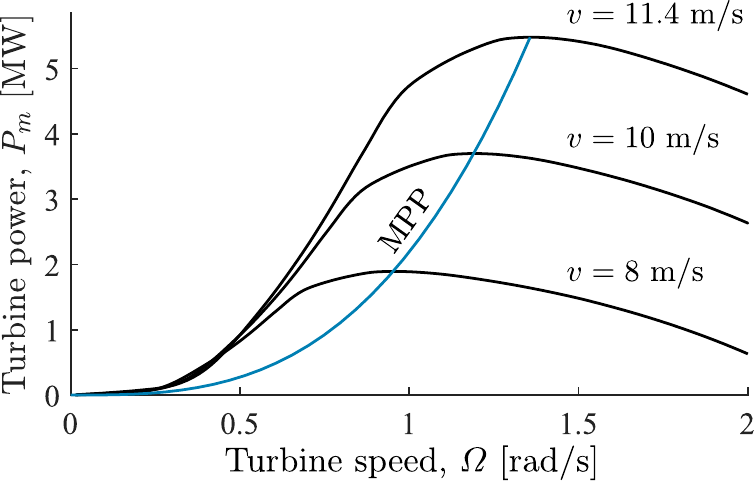}}}
        \caption{The NREL \SI{5}{\mega\watt} baseline WT model.}
  		\label{fig:NREL_model}
\end{figure*}

The  NREL WT model was developed to be representative of a typical utility-scale land- and sea-based WT~\cite{jonkmanDefinition5MWReference2009}, but did not envision the latest developments in the field e.g. synthetic inertia. To overcome this, we modify its control system so that the turbine can participate in FFR even when operated at the MPP. 
Here we give a brief overview of the simplified model depicted in \cref{fig:NREL_model}, which we use for the analysis and control design. 

The drivetrain and turbine properties are presented in \cref{tab:drive_train_prop}.
\begin{table}[t!]
\centering
\caption{Parameters of the NREL \SI{5}{\mega\watt} WT model.}
\label{tab:drive_train_prop}
\begin{tabular}{lrl}
\hline
Rated electric power & $P_\rated$: & \SI{5}{\mega\watt} \\
Torque rate limit & $\smash{\frac{d}{dt}\tau_\mathrm{set}}$:& $\pm$\SI{15}{\kilo\Newtonmeter\per\second} \\
Electric efficiency & $\eta$: & \SI{94.4}{\percent} \\
Rated speed  & $\varOmega_\rated$: & \SI{12.1}{rpm}  \\
Gearbox ratio &  $N$: & 97  \\
Inertia, high speed shaft & $J_e$: &  \SI{534.116}{\kgmeter\squared} \\
Inertia, low speed shaft & $J_m$: &  \SI{35444067}{\kgmeter\squared}
\\
Air density & $\rho$: &  \SI{1.225}{\kilogram\per\cubic\meter}
\\
Rotor radius & $r$: &  \SI{63}{\meter}
\\
Optimal tip speed ratio & $\lambda_\optimal$: &  7.5
\\
\hline
\end{tabular}
\end{table}
For the analysis, the stable drivetrain dynamics are ignored. Therefore, we do not differentiate between rotor speed and generator speed. Then, the total inertia with respect to the low speed shaft is
$
    J = J_m+N^2J_e = \text{\SI{40470000}{\kgmeter\squared}.}
$
The acceleration of the rotor is determined by the electric power output $P_e$ of the generator and the mechanical power input $P_m$ from the turbine. 
The main component of the  control system is the logic operator shown as $F_\textrm{logic}$ in \cref{fig:NREL_model}.
% The control logic, $F_\mathrm{logic}$, determines the operating mode of the turbine.
It controls $P_e$ by adjusting the electric torque set-point  $\tau_\mathrm{set}$ to  $\min (\tau_\mathrm{ref},\tau_\optimal)$, where the reference torque $\tau_\mathrm{ref}$ is given by the external power reference $P_\mathrm{ref}$ and $\tau_\optimal$ is the optimal torque read from the MPP lookup table. 
If the measured rotor speed $\hat \varOmega$ exceeds the rated speed $\varOmega_\rated$, then $F_\mathrm{logic}$ activates the pitch controller $F_\textrm{pitch}$ as to  stabilize the speed at $\varOmega=\varOmega_\rated$. This is done by increasing the pitch $\beta$, thereby reducing the aerodynamic power coefficient $c_p$ and thus the power $P_m$ extracted from the wind. Assuming operation below rated speed, then $\beta = 0$.  Thus, the power coefficient is only a function of the tip speed ratio $\lambda$. If $\tau_\mathrm{ref}\geq \tau_\optimal$, the turbine operates at the MPP curve shown in \cref{fig:wind_speed_curves}. 

The control structure of the original NREL model does not allow the WT to participate in FFR without curtailment. To amend this, the lookup table in \cref{fig:simplified_NREL_model} is replaced with a variable-speed feedback controller.

\subsection{Nordic 5-Machine (N5) Test System}
\label{sec:N5_windhydro}
\begin{figure}[t!]
		\centering
		\includegraphics[width=\linewidth]{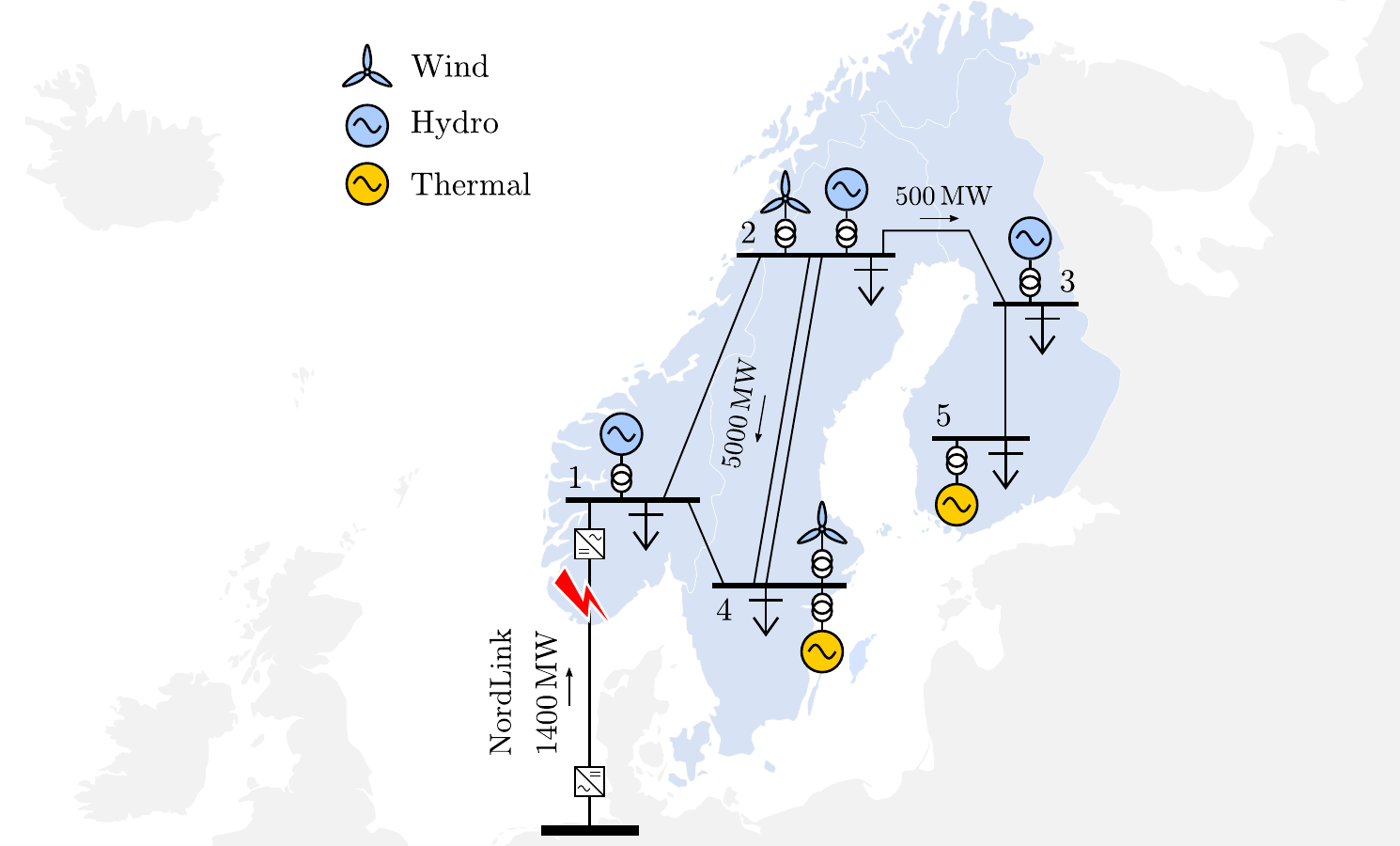}
		\caption{One-line diagram of the N5 test system.}
		\label{fig:Nordic_5}
\end{figure}

Consider the N5 test system shown in \cref{fig:Nordic_5}; adapted from the empirically validated 3-machine model presented in~\cite{saarinenFullscaleTestModelling2016}. The system is fictitious but has dynamic properties similar to those of the Nordic synchronous grid. The test system is  implemented in Simulink Simscape Electrical~\cite{hydro-quebecSimscapeElectricalReference2020_manual}.
Loads, synchronous machines, and WTs are lumped up into a single large unit at each bus. The hydro and thermal units are modeled as 16\textsuperscript{th} order salient-pole and round rotor machines, respectively. 
Assuming that inverters are operated within the allowed limits and have high enough bandwidth so that they have no significant impact on the studied frequency dynamics, we assume that all inverters can be modeled as controllable power loads. The WTs, modified to participate in FFR, are of 8\textsuperscript{th} order.\footnote{
The full model, and test cases presented in this work are available at the {GitHub repository \url{https://github.com/joakimbjork/Nordic5}.}}

The system's kinetic energy varies greatly over the year, as the amount of synchronous generation connected to the grid depends on the demand~\cite{entso-eFastFrequencyReserve2019}.
For this analysis, we consider a low inertia scenario with $W_\mathrm{kin} = \text{\SI{110}{\giga\wattsecond}}$ distributed according to \cref{tab:Nordic5}. Loads are modeled as constant power loads with a combined frequency dependency of \SI{400}{\mega\watt\per\hertz}. 

\begin{table}[t!]
\centering
\caption{Machine parameters for the \SI{110}{\giga\wattsecond} test case. 
Time constants and distribution of FCR are based on the case study in~\cite{saarinenFullscaleTestModelling2016}.
}
\label{tab:Nordic5}
\begin{tabular}{cc|cc|ccc}
\hline
  Bus & $W_\mathrm{kin}$ [\si{\giga\wattsecond}] & $P_\mathrm{gen}$ [\si{\mega\watt}] & FCR [\si{\percent}] &
  $T_y$ & $T_\mathrm{w}$ & $g_0$ \\
 \hline
1 & 34 & \SI{9000} & 60 & 0.2 & 0.7 & 0.8 
\\
2 & 22.5 & \SI{6000} & 30 & 0.2 & 1.4 & 0.8 
\\
3 & 7.5 & \SI{2000} & 10 & 0.2 & 1.4 & 0.8 
\\
4 & 33 & \SI{5000} & -- & -- & -- & -- 
\\
5 & 13 & \SI{2000} & -- & -- & -- & -- 
\\
\hline
\end{tabular}
\end{table}

\begin{figure}[t!]
		\centering
		\includegraphics[width=\linewidth]{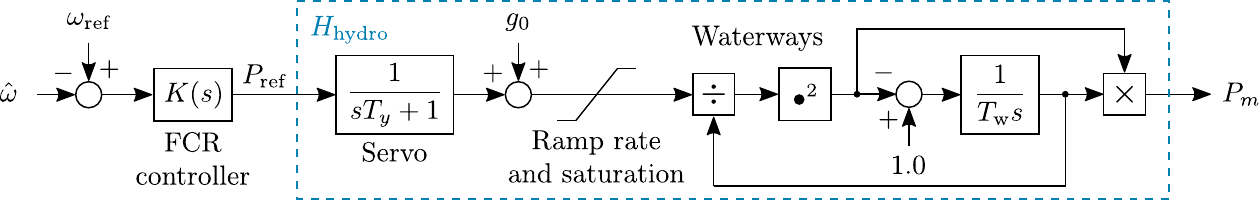}
		\caption{Block diagram of the hydro turbine and governor model.}
		\label{fig:hydro_block}
\end{figure}

The FCR control is designed to ensure that the center of inertia (COI) frequency
\begin{equation}
    \omega_\mathrm{COI} = \frac{\sum_{i=1}^5{ W_{\textrm{kin},i} \omega_i}}{\sum_{i=1}^5 W_{\textrm{kin},i}}
\end{equation}
is kept within allowed dynamic bounds. Here, $\omega_i$ notates the synchronous rotor speed and $W_{\textrm{kin},i}$ the total kinetic energy of machines in area $i$. In the Nordic synchronous grid, FCR is primarily provided by hydro.
The hydro governor implemented in this work is an adaption of the model available in the Simulink Simscape Electrical library~\cite{hydro-quebecSimscapeElectricalReference2020_manual}. It has been modified to allow a general linear FCR controller, $K(s)$, instead of the fixed PID/droop control structure, as shown in \cref{fig:hydro_block}. The servo rate limit is set to the default $\pm\text{\SI{0.1}{\perunit\per\second}}$. The nonlinear second-order model is useful for large-signal time-domain simulations.
For the linear analysis, the turbine is modeled as
\begin{equation}
\label{eq:Hhydro}
    H_\textrm{hydro}(s) = 
     2\frac{ z - s}{s + 2z}  
    \frac{1}{sT_y + 1} 
    , \quad z = \frac{1}{g_0 T_\mathrm{w}}
\end{equation}
where $T_y$ is the servo time constant, $g_0$ the initial gate opening, $T_\mathrm{w}$ the water time constant,  $\hat \omega$ the locally measured frequency, and $\omega_\mathrm{ref}$ the frequency reference. 

\subsection{The FCR and FFR Control Problem}
\label{sec:theFCR_control_problem}
In this work, we study how FFR from wind can be coordinated with FCR from hydro, to fulfill an ``ideal" FCR design goal that cannot be achieved individually by the two resources. To do this in a methodical manner, we derive a linear model, similar to the well-known hydro model \eqref{eq:Hhydro}, that captures the shortcomings of wind-FFR.

To specify a desired FCR response, we use the FCR for disturbance (FCR-D) specifications in the Nordic synchronous grid. 
The FCR-D is used to contain the frequency outside normal operation. 
Following a rapid frequency fall from 49.9 to \SI{49.9}{\hertz}, the reserves should be \SI{50}{\percent} activated within \SI{5}{\second} and fully activated in \SI{30}{\second}.
Following larger disturbances the maximum instantaneous frequency deviation (the nadir) should be limited to \SI{1}{\hertz}~\cite{entso-eNordicSynchronousArea2018}.

Let the FCR-D design target take the form 
\begin{equation}
\label{eq:asmFCRD}
%    F_\FCR(s) := R_\FCR \frac{1}{s T_\FCR + 1} \ [\si{\mega\watt \per \hertz}].
    F_\mathrm{FCR}(s) =  R_\mathrm{FCR} \frac{6.5s+1}{(2s+1)(17s+1)}.
\end{equation}
% \end{asm}
Consider the dimensioning fault to be the instant disconnection of the NordLink dc cable~\cite{NordLink_manual} importing \SI{1400}{\mega\watt} from Germany into Norway as shown in \cref{fig:Nordic_5}.
Choosing $R_\mathrm{FCR} = \text{\SI{3100}{\mega\watt\per\hertz}}$, the post-fault system stabilizes at \SI{49.5}{\hertz}. The second-order filter in \cref{eq:asmFCRD} is tuned so that the FCR-D requirements are fulfilled, while also avoiding an overshoot and a second frequency dip when the frequency is restored. {For a description of how the FCR-D design target can be selected, see the Appendix.}

\begin{figure}[t!]
    \captionsetup[subfloat]{farskip=0pt}
    \centering
    \subfloat[\label{fig:Nordic5_hydro_1} FCR response.]
    {{\includegraphics[scale=0.52]{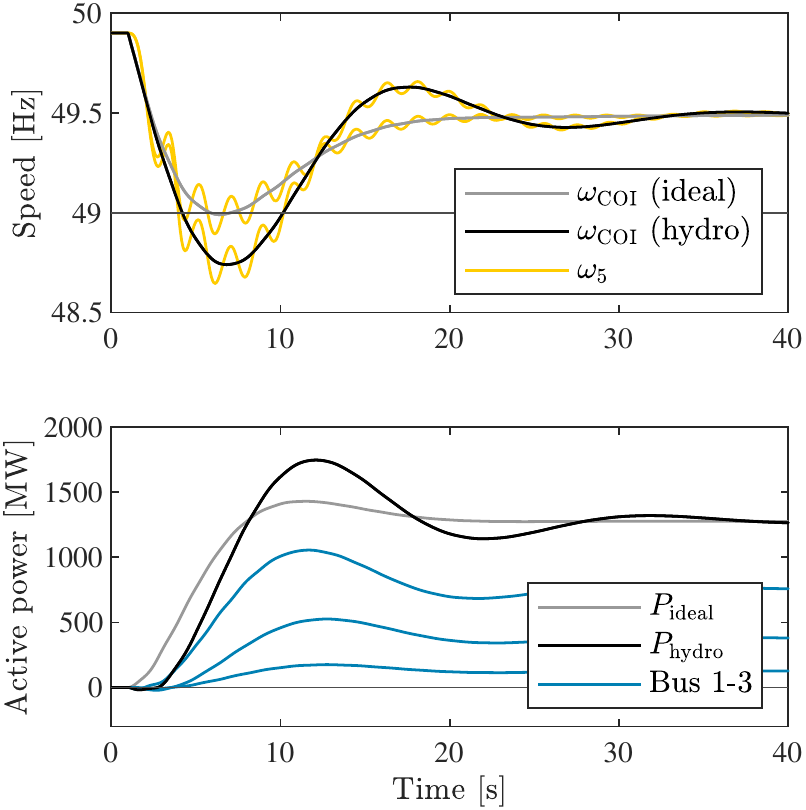}}}
    \hfill
    \subfloat[\label{fig:Nordic5_hydro_zoom} Zoom in on hydro FCR response.]
    {{\includegraphics[scale=0.52]{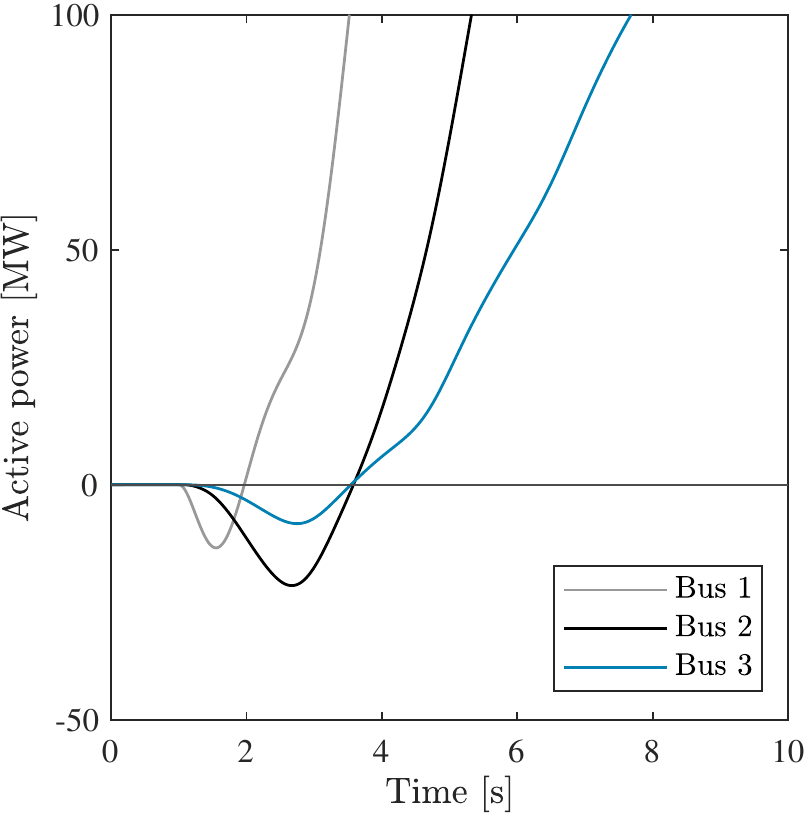}}    }
    \caption{System response to a \SI{1400}{\mega\watt} fault with hydro FCR.}
    \label{fig:Nordic5_hydro}
\end{figure}

To allocate FCR between the participating devices, we use the model matching design procedure developed in~\cite{bjorkDynamicVirtualPowerunpublished,bjorkFundamentalControlPerformance2021}. Assume that the power injection, $P_i$, of the frequency depended actuator $i\in\{1,\ldots,n\}$ can be described by the linear model
\begin{equation}
    P_i = H_i(s)K_i(s)(\omega_\mathrm{ref}-\hat \omega).
\end{equation}
The FCR or FFR controllers $K_i(s)$ are designed by selecting dynamic participation factors $c_i(s)$ that take into account the dynamic strengths and weaknesses of participating devices, and to align with economic considerations. 
We say that perfect matching is achieved if
\begin{equation}
\label{eq:perfect_match}
    \sum_{i=1}^{n}c_i(s) = 1, \quad \forall s \in \complex.
\end{equation}
Choosing $K_i(s) =c_i(s)F_\FCR(s)/H_i(s)$, then the FCR-D requirements are fulfilled with $P_\mathrm{ideal} = \sum_{i=1}^{n} P_i$, as shown in \cref{fig:Nordic5_hydro_1}.

In the Nordic grid, FCR is almost exclusively provided by hydropower. For internal stability, $c_i(s)$ need to include all NMP zeros in the plant $H_i(s)$~\cite{bjorkDynamicVirtualPowerunpublished,bjorkFundamentalControlPerformance2021}. With only hydro units \cref{eq:Hhydro}, we can therefore not achieve perfect matching \cref{eq:perfect_match}. With total FCR ($P_\mathrm{hydro}$) delivered by the hydro units at buses 1, 2, and 3, the FCR-D requirements are no longer fulfilled using the design target \cref{eq:asmFCRD}. 
This is because the initial surge of power from the hydro turbines are in the opposite direction of the gate opening change, as seen in \cref{fig:Nordic5_hydro_zoom}.
As the gate opens, the pressure over the turbine falls before the water, due to the inertia in the water column, has time to accelerate to a new steady-value~\cite{kundurPowerSystemStability1994}. 
This behaviour is characterized by the NMP zero in  \cref{eq:Hhydro}. The implemented hydro model also captures nonlinear ramp rate and saturation constraints. However, with hydro-FCR resources distributed according to \cref{tab:Nordic5}, absolute capacity and rate limiters are not a problem. The dynamic limitations due to NMP zeros will be more relevant.

The design target \cref{eq:asmFCRD} can be modified so that the FCR-D requirements are fulfilled even if FCR are delivered by hydro governors. However, due to the bandwidth limitations imposed by the NMP zeroes, this is not a good solution since this reduces the closed-loop stability margins~\cite{agneholmFCRDDesignRequirements2019}. Because of this, the Nordic SOs have developed a new market for FFR~\cite{entso-eFastFrequencyReserve2019}.

\section{Design of a New Variable-Speed WT Model} \label{sec:WTmodel}
In this  section, we design a new model-based variable-speed feedback control, allowing WTs to participate in FFR without curtailment. The control scheme is similar to~\cite{morrenWindTurbinesEmulating2006}, but is designed so that the dynamics relevant for FFR can be described by a first-order model.  The result is a first-order linear model, similar to the commonly accepted hydro governor model \eqref{eq:Hhydro}.
The simplified linear representation conveniently enables the coordination with other power sources using \cref{eq:perfect_match}.
The simplified model is compared to the detailed nonlinear model in a simulation study, validating that the linearized model captures dynamics relevant for FFR control design.

The dynamic properties of the WT vary greatly with the wind speed. Because of this, it is convenient to express the dynamics in terms of the normalized speed ratio
\begin{equation}
\label{eq:norm_speed_ratio}
    x:= {\varOmega}/{\varOmega_\MPP} = {\lambda}/{\lambda_\optimal},
\end{equation}
where $\varOmega_\MPP$ is the MPP turbine speed.

\subsection{Open-Loop Stable and Unstable Operating Modes}
The open-loop characteristic of the WT are described by the nonlinear differential equation 
\begin{equation}
    \frac{d}{dt}\varOmega  = \frac{\varOmega R(\varOmega) - P_e}{M(\varOmega)}.
\end{equation}
Linearizing around the operating point $\varOmega = \varOmega_0$, $P_e = P_0$, and $x = x_0 = \varOmega_0/\varOmega_\MPP$, with some abuse of notation, we get the
transfer function
\begin{equation}
\label{eq:open-loop_nonlinear}
    \varOmega = \frac{-1}{s M(\varOmega_0) - R(\varOmega_0)} P_e,
\end{equation}
where the inertia 
\begin{equation}
    \label{eq:M}
     M(\varOmega_0) = J {\varOmega_\MPP }x_0
\end{equation}
and the power coefficient derivative
\begin{equation}
\label{eq:def_R}
    R(\varOmega_0) = 
    \left.\frac{\partial}{\partial \varOmega} P_m \right|_{\varOmega=\varOmega_0}
    % = 
    %  P_\wind \frac{\partial}{\partial \varOmega} c_p 
    =
    \left.\frac{P_\wind}{\varOmega_\MPP}  \frac{\partial}{\partial x} c_p\right|_{x=x_0}.
\end{equation}

\begin{figure}[t!]
    \captionsetup[subfloat]{farskip=0pt}
    \centering
    \subfloat[\label{fig:wind_Pm_w_slope}Power coefficient.]
    {{\includegraphics[width=0.49\linewidth]{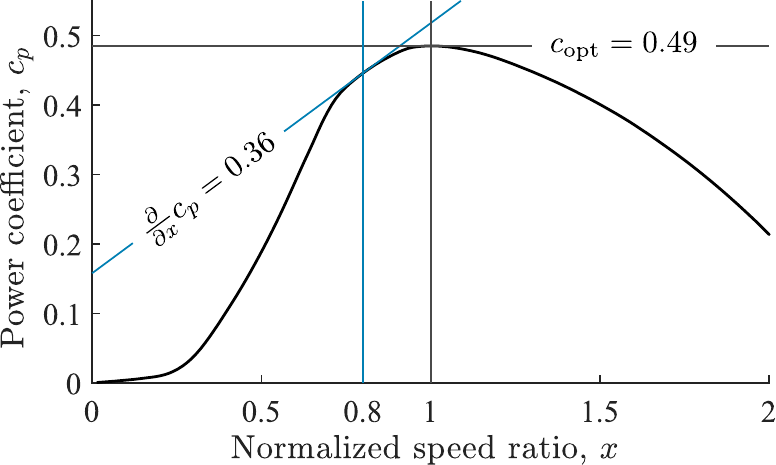}}}
    \hfill
    \subfloat[\label{fig:wind_Pm_w_ratio}Derivative of power coefficient.]
    {{\includegraphics[width=0.49\linewidth]{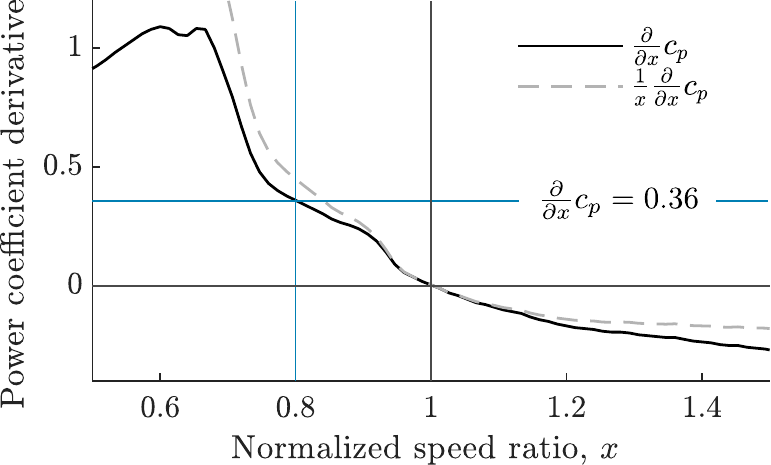}}}
    \caption{Power efficiency coefficient of the NREL \SI{5}{\mega\watt} turbine with $\beta=0$.}
    \label{fig:wind_speed_normalized}
\end{figure}

Due to the nonlinear power--speed characteristics in \cref{fig:wind_speed_normalized}, the WT presents both stable and unstable operating modes.

\subsubsection{Stable Operating Mode}
Forcing the turbine to accelerate above the optimal speed reduces the mechanical output power, which automatically reduces the acceleration. This, intuitively results in a stable mode of operation which can be verified by \eqref{eq:def_R}. As seen in \cref{fig:wind_Pm_w_ratio}, $\frac{\partial}{\partial x} c_p < 0$ for an operation above optimal speed. Thus, the open-loop system \eqref{eq:open-loop_nonlinear}  has a single pole in the left-half complex plane 

\begin{equation}
    s = R(\varOmega_0)/M(\varOmega_0) < 0, \quad \text{for} \quad x_0 > 1
\end{equation}
and is therefore stable.

\subsubsection{Unstable Operating Mode}
\label{sec:unstable_op_mode}
Forcing the turbine to decelerate below optimal speed reduces the mechanical output power, which is automatically enhanced as it continues to reduce the mechanical power. This results in an unstable operating mode. Again, from \eqref{eq:open-loop_nonlinear} we see that the open-loop system will have an unstable pole at
\begin{equation}
\label{eq:unstable_pole}
    s  = R(\varOmega_0)/M(\varOmega_0)> 0, \quad \text{for} \quad x_0 < 1.
\end{equation}

If we assume that the turbine is controlled, such that the normalized speed ratio is bounded from below by $\underline{x} > 0$, then $M(\varOmega_0)  > J {\varOmega_\MPP } \, \underline{x}$. Moreover, as seen in \cref{fig:wind_Pm_w_ratio}, $\frac{\partial}{\partial x} c_p > 0$. 
Consequently, the unstable open-loop pole \eqref{eq:unstable_pole} is bounded from above by
\begin{equation}
\label{eq:pole_bound}
     s =
     \frac{ P_\wind}{J \varOmega_\MPP^2} 
     \frac{1}{x_0} \left.
     \frac{\partial}{\partial x} c_p  \right|_{x=x_0}
     <
     \frac{ P_\wind}{J \varOmega_\MPP^2} 
     \frac{1}{\underline x} 
     \left. \frac{\partial}{\partial x} c_p \right|_{x = \underline{x}}.
    %\overset{\underline{x} = 0.8}{< }
  % \frac{ P_\wind}{J \varOmega_\MPP^2} 
   % \frac{0.36}{0.8} 
\end{equation}

\begin{rem}[Scalability]
\label{rem:scalability}
% using \cref{eq:def_Pwind,eq:def_lambda}
Note that 
    \begin{equation}
        \frac{P_\wind}{J} \frac{1}{\varOmega_\MPP^2 }
        =
         \frac{\rho \pi r^2 v^3}{2J}
        \frac{r^2}{\lambda_\optimal^2v^2} 
        % \frac{1}{\varOmega_\MPP^2 }
        %  \frac{\rho \pi r^2 v^3}{2 J \varOmega_\MPP^2} 
         =
          \frac{\rho \pi r^2 }{2J} \frac{r^2}{\lambda_\optimal^2} v
        .
    \end{equation}
% The second factor in \eqref{eq:pole_bound} is a nonlinear unit-less variable that depends on the normalized speed ratio \eqref{eq:norm_speed_ratio} as shown in \cref{fig:wind_Pm_w_ratio}. 
Thus, the open-loop pole \eqref{eq:pole_bound}
%\begin{equation}
%\label{eq:pole_equation}
%    s= 
%    \underbrace{\frac{ \pi r^2}{J}}_{
%    \substack{\textrm{size/inertia} \\ \textrm{constant} }   } 
%    \frac{\rho }{2} 
%    \underbrace{\frac{r^2}{\lambda_\optimal^2} }_{
%     \substack{\textrm{aerodynamic} \\ \textrm{constant} } }
%     v \frac{1}{x_0} \left.\frac{\partial}{\partial x} c_p\right|_{x=x_0} = C  v \frac{1}{x_0} \left. \frac{\partial}{\partial x} c_p\right|_{x=x_0}
%\end{equation}
\begin{equation}
	\label{eq:pole_equation}
	s= 
	\frac{ \pi r^2}{J}
	\frac{r^2}{\lambda_\optimal^2} 
	\frac{\rho }{2} 	
	v \frac{1}{x_0} \left.\frac{\partial}{\partial x} c_p\right|_{x=x_0} = C  v \frac{1}{x_0} \left. \frac{\partial}{\partial x} c_p\right|_{x=x_0}
\end{equation}
varies linearly with the wind speed $v$. The first coefficient in \eqref{eq:pole_equation}, ${\pi r^2}/{J}$, is proportional to size/inertia; whereas the second coefficient, ${r^2}/{\lambda_\optimal^2}$, can be seen as an aerodynamic constant for the WT. The constant $C$ should be fairly consistent for all WT models. Thus, the location of the unstable pole should be similar for WTs of various ratings.
\end{rem}

\subsection{Variable-Speed Feedback Controller}
\label{sec:stab:ctrler}
\begin{figure}[t!]
		\centering
		\includegraphics[width=\linewidth]{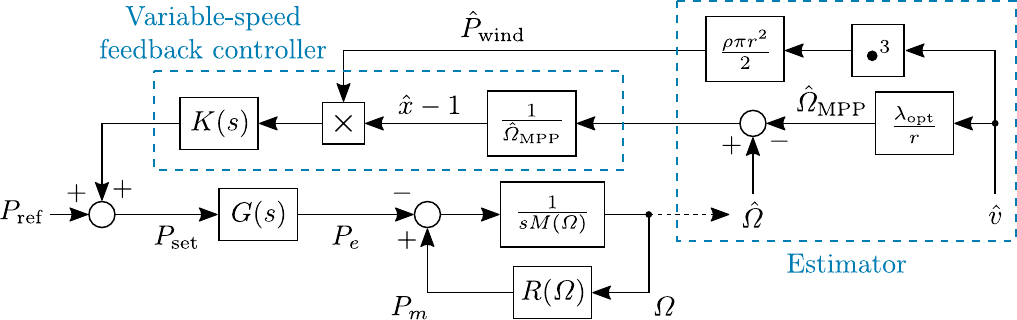}
		\caption{Block diagram of WT with variable-speed feedback control.}
		\label{fig:minimal_wind_model}
\end{figure}

To operate the WT in the unstable region, we implement a stabilizing variable-speed feedback controller.
The closed-loop system in \cref{fig:minimal_wind_model} from reference to output is 
\begin{equation}
\label{eq:closed_loop_exact}
  P_e = G(s)\frac{s M(\varOmega_0)-R(\varOmega_0)}{s M(\varOmega_0)+ G(s)\hat{K}(s) -R(\varOmega_0) } P_\textrm{ref}.
\end{equation}
Note that the closed-loop system has a NMP zero at the location of the unstable open-loop pole that is unaffected by the generator and measurement dynamics $G(\jomega)$ and the controller design. 

Let $K(s)$ be a linear controller and let
\begin{equation}
\label{eq:Kstab_variable}
    \hat K(s) = K(s) {\hat{P}_\wind}/\hat \varOmega_\MPP
\end{equation}
where the notation $\hat{P}_\wind$ and $\hat \varOmega_\MPP$ indicate estimates of ${P}_\wind$ and $\varOmega_\MPP$ respectively, e.g., from wind speed measurements as shown in \cref{fig:minimal_wind_model}.
The purpose of $\hat{K}(s)$ is to stabilize the WT. {For the NREL turbine, $|G(\jomega)| \approx 1$ in the frequency range of interest. Therefore, a proportional feedback controller
$    \hat{K}(s) = k {\hat P_\wind}/{\hat \varOmega_\MPP}  $
will be sufficient to stabilize the WT. This controller will move the unstable pole \eqref{eq:unstable_pole} of \eqref{eq:closed_loop_exact} into the left half-plane (LHP).}
Neglecting the generator and measurement dynamics, the WT is described by the transfer functions
\begin{equation}
\label{eq:non_linear_plant}
    P_e = \frac{s-{z}}{s+{p}} P_\textrm{ref} \quad \text{and} \quad
    \varOmega = \frac{-1}{J\varOmega_\MPP(s+{p})} P_\textrm{ref}.
\end{equation}

\subsection{Linearization Capturing Undesirable NMP Characteristics}
\label{sec:linearized_system}
Assuming operation at a normalized speed $\underline{x} <x<1$, $\underline x = 0.8$.
The unstable open-loop pole \eqref{eq:pole_bound}, and consequently the NMP closed-loop zero
\begin{equation}
\label{eq:zero_location}
    z = C  v \frac{1}{x_0} \left. \frac{\partial}{\partial x} c_p\right|_{x=x_0}   \leq \bar{z} = C v \frac{1}{\underline x} 
    \left. \frac{\partial}{\partial x} c_p \right|_{x = \underline x}
     .
\end{equation}
With a proportional feedback controller 
\begin{equation}
\label{eq:stablized_pole}
    p = k \frac{\hat P_\wind}{\hat \varOmega_\MPP} \frac{1}{J \varOmega_\MPP x_0} - z 
    \approx 
    C v \frac{1}{x_0}\left(k -  \left.\frac{\partial}{\partial x} c_p\right|_{x=x_0}\right)
\end{equation}
with equality if $\hat P_\wind = P_\wind $ and $\hat \varOmega_\MPP = \varOmega_\MPP$. Note that since the zero is bounded from above, the stabilized pole \eqref{eq:stablized_pole} is bounded from below by
\begin{equation}
\label{eq:stablized_pole_location}
p \geq
 \underline p 
    = 
    C v \frac{1}{\underline x}\left(k - \left. \frac{\partial}{\partial x} c_p \right|_{x = \underline x}
    \right). 
\end{equation}

For the analysis, \cref{eq:non_linear_plant} are linearized by setting $z$ and $p$  to their bounds $\bar{z}$ and $\underline{p}$, respectively. {This will overestimate the decline in output power since the steady-state gain
\begin{equation}
\label{eq:decline}
     -{\bar{z}}/{\underline{p}} \leq -{z}/{p} \leq 0.
\end{equation}
}
A good starting point is to select $k$ so that the unstable pole \eqref{eq:pole_bound} is reflected into the LHP, in which case $\underline{p} = \bar{z}$. 
Inserting values from the NREL \SI{5}{\mega\watt} WT into \eqref{eq:pole_equation}, then
\begin{equation}
\begin{aligned}
\label{eq:settings}
    C = 0.013, \quad   \left. \frac{\partial}{\partial x} c_p \right|_{x = 0.8} = 0.36, \quad 
    k = 2 \times 0.36,
    \end{aligned}
\end{equation}
and the transfer function from $P_\textrm{ref}$ to $P_e$
\begin{equation}
\label{eq:linear_plant_approx}
    H_\textrm{wind}(s) = \frac{s-\bar{z}}{s+\bar{z}}, \quad \bar{z} =  5.8 v\cdot 10^{-3}.
\end{equation}
To be useful in the coordinated FCR and FFR control design, we want the simplified linear model \cref{eq:linear_plant_approx} to be a "worst-case" realization of \eqref{eq:non_linear_plant} in that it underestimates the WT's power output. This is achieved by choosing the upper bound \cref{eq:zero_location} for the zero and lower bound \cref{eq:stablized_pole_location} for the pole since this overestimate the decline in output power \eqref{eq:decline}.

\begin{rem}[MPP Estimation Bias]
Although often measured, wind speed measurements may be an inaccurate indicator of the actual wind energy captured by the rotor blades. It may therefore give a biased estimation of the MPP. One way around this is to instead use feedback from the rotor or generator speed. This is the design most often used by conventional MPP tracking controllers~\cite{morrenWindTurbinesEmulating2006,ochoaFastfrequencyResponseProvided2017,ullahTemporaryPrimaryFrequency2008,liCoordinatedControlStrategies2017,leeReleasableKineticEnergyBased2016,zhaoFastFrequencySupport2020,wilches-bernalFundamentalStudyApplying2016,jonkmanDefinition5MWReference2009}. The transfer function \cref{eq:closed_loop_exact} of a WT with a stabilizing controller based on speed measurements will however be more nonlinear, making the linear analysis less transparent. As a proof of concept, this paper therefore considers a MPP estimator based on local wind speed measurements. Note however that the NMP zero dynamics of \cref{eq:closed_loop_exact} are independent of the controller design. The fundamental control limitations captured by the simplified model will therefore be the same, regardless of the how the stabilizing control is implemented.
\end{rem}

\subsection{Validating the Properties of the Simplified Linear WT Model}
The purpose of the simplified model \eqref{eq:linear_plant_approx} is to capture the dynamics most relevant for FFR. Because of this, it is designed to underestimate the electric power output of the WT. In addition, the model also need to capture the dynamics that are relevant for safe operation, with minimal conservatism. Since the most important property in this context is the normalized speed ratio \eqref{eq:norm_speed_ratio}, the simplified linear model is designed to overestimate the decline in turbine speed. To ensure stability of the WT, regardless of the power reference input, we also implement a low speed protection mechanism.
To show that the simplified model \eqref{eq:linear_plant_approx} possess the desired properties, we here compare it with the detailed nonlinear modified NREL WT model. {First we show how the WT compares to the simplified model under ideal conditions with constant wind speeds. Then we show a more realistic scenario with varying wind speeds.}
Model properties such as pole and zero bounds for various wind speeds and controller gains are shown in \cref{tab:linear_sim_cases}.

\subsubsection{Different Wind Speeds}
Let $k = 0.72$ so that the WT is stable for operation with a normalized speed ratio above $ \underline x = 0.8$. 
Consider operation at wind speeds 8 and \SI{10}{\meter\per\second}. 

\begin{figure}[t!]
    \captionsetup[subfloat]{farskip=0pt}
    \centering
    \subfloat[\label{fig:openloop_wind_speeds_SI}SI-units.]
    {{\includegraphics[width=0.49\linewidth]{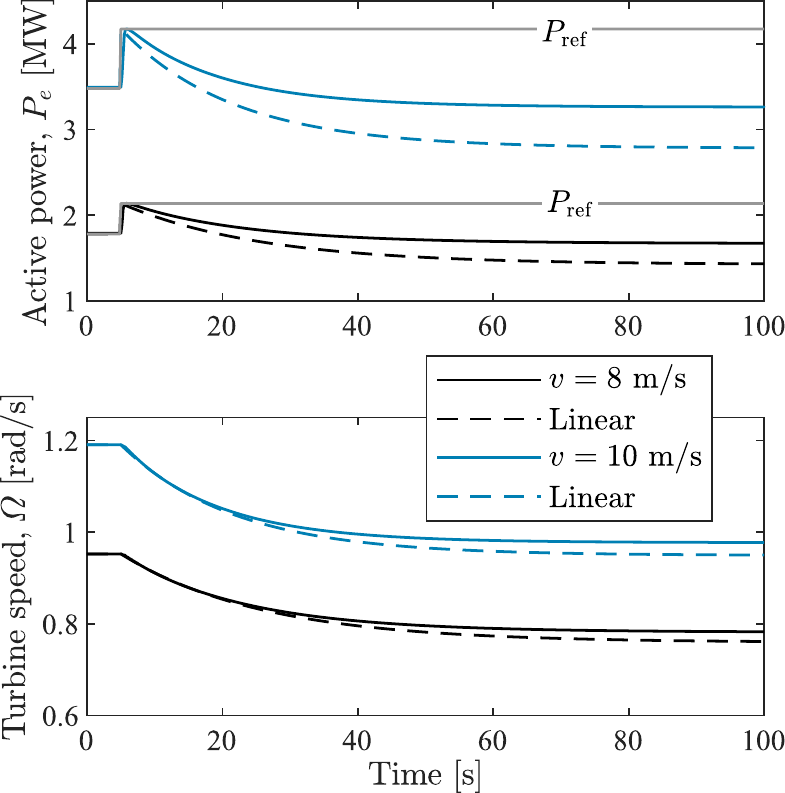}}}
    \hfill
    \subfloat[\label{fig:openloop_wind_speeds_normalized}Normalized.]
    {{\includegraphics[width=0.49\linewidth]{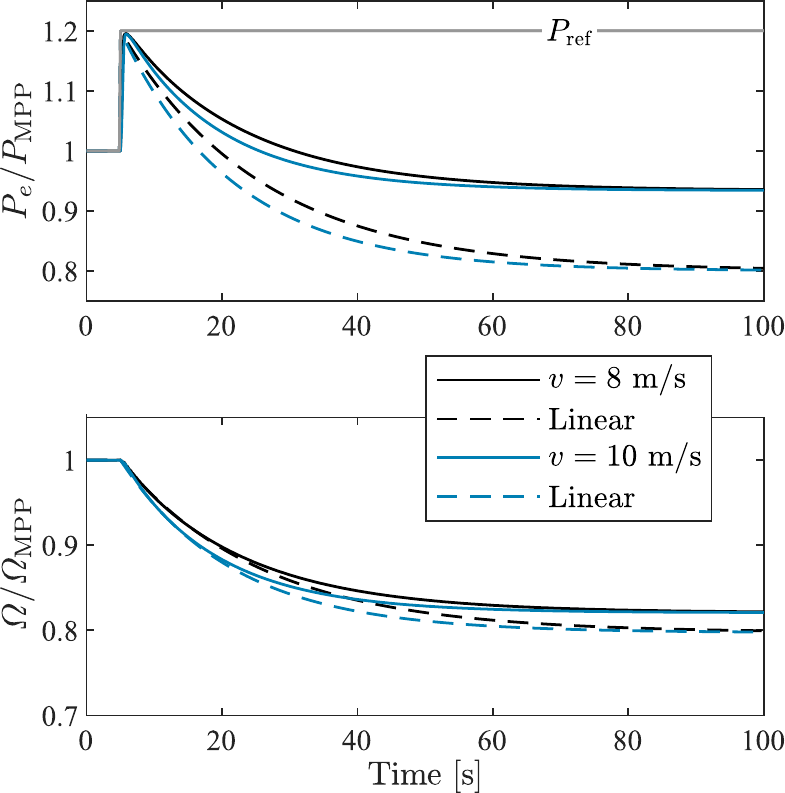}}}
    \caption{Power output and rotor speed with feedback gain $k=0.72$.}
    \label{fig:openloop_wind_speeds}
\end{figure}

As shown in \cref{fig:openloop_wind_speeds}, after a $+\text{\SI{20}{\percent}}$ reference step, the active electric power output is initially increased above the MPP. Consequently, the rotor speed decreases, which reduces the aerodynamic efficiency of the WT according to \cref{fig:wind_speed_curves}. Therefore, the increased power output cannot be sustained for long and eventually falls below the initial MPP output.

The location of the open-loop pole \eqref{eq:pole_equation}  and consequently the closed-loop zero \eqref{eq:zero_location} varies linearly with the wind speed. However, since the variable-speed controller \eqref{eq:Kstab_variable} uses feed forward of the estimated wind energy $\hat P_\wind$, it will be equally efficient for different wind speeds, as seen in \cref{fig:openloop_wind_speeds_normalized}.

Note that the linear model \eqref{eq:linear_plant_approx} is conservative in the sense that it overestimates the size of the nonlinear NMP zero \eqref{eq:zero_location} which varies with the power coefficient derivative \cref{fig:wind_Pm_w_ratio}. Thus the linear model overestimates the decline in power from the turbine. Similarly it overestimates the decline in turbine speed since the pole \eqref{eq:stablized_pole_location} is underestimated. 

\begin{table}[t!]
\centering
\caption{Model properties for the step response simulations.}
\label{tab:linear_sim_cases}
\begin{tabular}{cc|cc|cc}
\hline
  $k$ & $v$ [\si{\meter\per\second}] & $P_\MPP$ [\si{\mega\watt}] & $\varOmega_\MPP$  [\si{\radian\per\second}]  & $\bar{z}$ [\si{\radian\per\second}] & $\underline{p}$ [\si{\radian\per\second}] \\
 \hline
  0.72 & 8 & 1.8 & 0.95  & 0.048& 0.048 
 \\
   0.72 & 10 & 3.5 & 1.19 & 0.060& 0.060
  \\
   1.08 & 8 & 1.8 & 0.95 & 0.048& 0.096
  \\
\hline
\end{tabular}
\end{table}

\subsubsection{Different Feedback Gains}
The choice of feedback gain affects the size of of the stabilized pole $\underline{p}$ as shown in \cref{tab:linear_sim_cases}. In \cref{fig:increased_gain} we see that increasing the feedback gain makes the output decline faster. Consequently, the rotor speed and output power stabilizes at a higher level.

\subsubsection{Low Speed Protection}
\begin{figure}[tb!]
    \captionsetup[subfloat]{farskip=0pt}
    \centering
    \subfloat[\label{fig:increased_gain}Different feedback gains.]
    {{\includegraphics[width=0.49\linewidth]{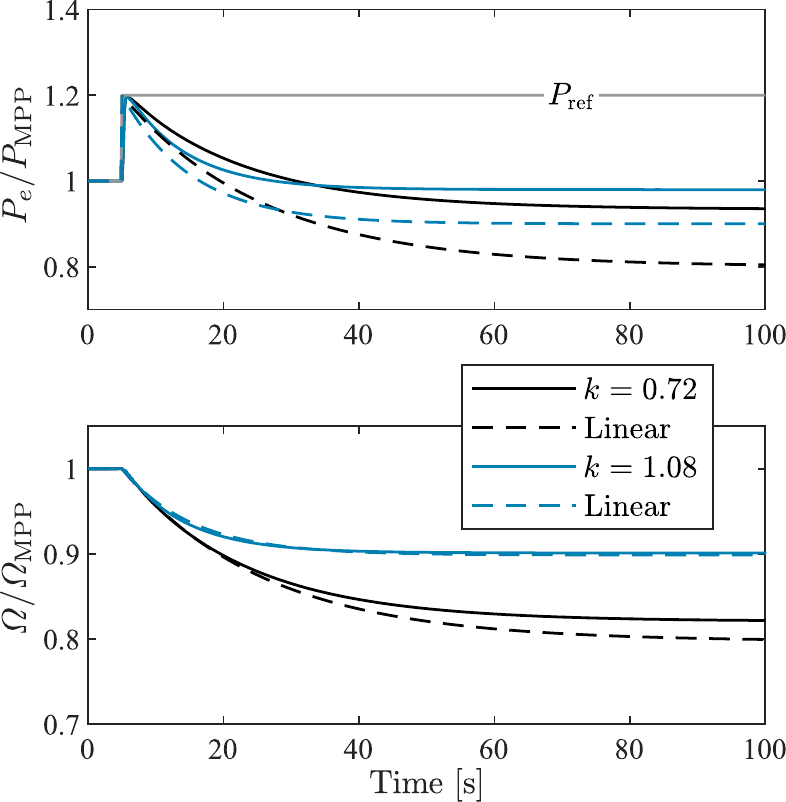}}}
    \hfill
    \subfloat[\label{fig:increased_step}Low speed protection.]
    {{\includegraphics[width=0.49\linewidth]{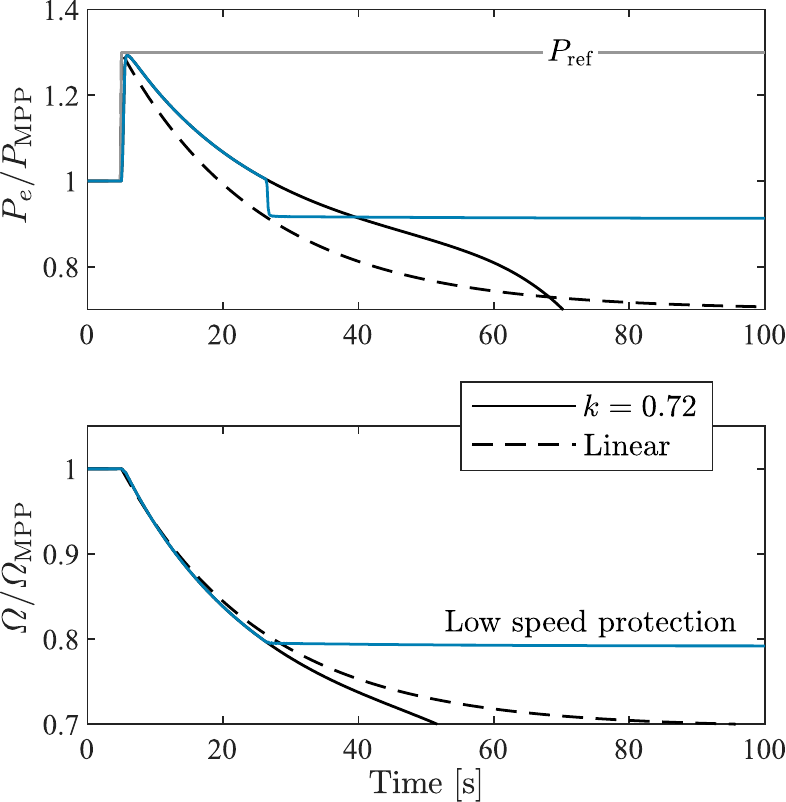}}}
    \caption{Power output and rotor speed at \SI{8}{\meter\per\second} wind speed.}
    \label{fig:openloop_wind_speeds2}
\end{figure}

For stability, it is important that the normalized speed ratio does not decline below $x=0.8$, since for $x< 0.8$, the power efficient derivative changes rapidly as shown in \cref{fig:wind_Pm_w_ratio}. With the stabilizing gain $k = 0.72$ we have designed the turbine to safely operate at normalized speeds $x> 0.8$. As shown in \cref{fig:increased_gain}, a $+\text{\SI{20}{\percent}}$ reference step almost reaches the speed limit. Increasing the step size to $+\text{\SI{30}{\percent}}$ makes the turbine decelerate to much so that it becomes unstable, as seen in \cref{fig:increased_step}.

Stability can be ensured by increasing the feedback gain. However, when controlling the WT to provide FFR, the focus is on the output power, $P_e$, not on the internal operation of the WT. Stable poles can then very well be cancelled by a FFR controller, as we will see in \cref{sec:DVPP}. The location of the stable WT pole is therefore not critical in closed-loop FFR control, as long as stability is ensured.
To hinder accidental destabilization of the WT,  a low speed safety measure is implemented, that disables the FFR controller and halts the deceleration when the normalized speed fall below $x=0.8$. Let 
\begin{equation}
\label{eq:low-speed-prot}
     P_\textrm{set}=\left\{ 
    \begin{aligned}
    &
     P' = P_\textrm{ref}   + \hat P_\wind {k}(\hat x - 1)
    ,
    \\
     & \min 
    \Big( P_{0.8} \big(1 - 100 (\hat x - 0.8)^2 \big) ,  P' \Big),   
    \end{aligned}
    \
    \begin{aligned}
    & \hat x>0.8
    \\
    & \hat x<0.8
    \end{aligned}
    \right.
\end{equation}
where  $P_{0.8} = P_m(x=0.8)$.
As shown in  \cref{fig:increased_step,fig:Nordic5_x2} this stabilizes the WT, bringing the turbine speed back to $x \approx 0.8$.

\subsubsection{Varying Wind Speed} 
\label{sec:vary_speed}
\begin{figure}[tb!]
    \captionsetup[subfloat]{farskip=0pt}
    \centering
    \subfloat[\label{fig:variable_wind}Recorded wind speed~\cite{syslab}.]
    {{\includegraphics[width=0.49\linewidth]{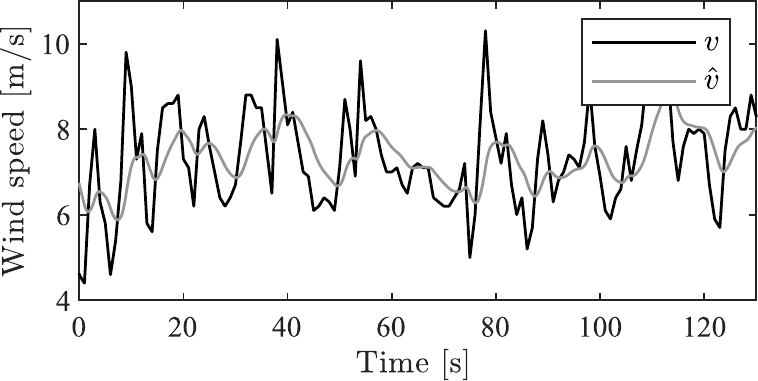}}}
    \hfill
    \subfloat[\label{fig:variable_track}Power injection.]
    {{\includegraphics[width=0.49\linewidth]{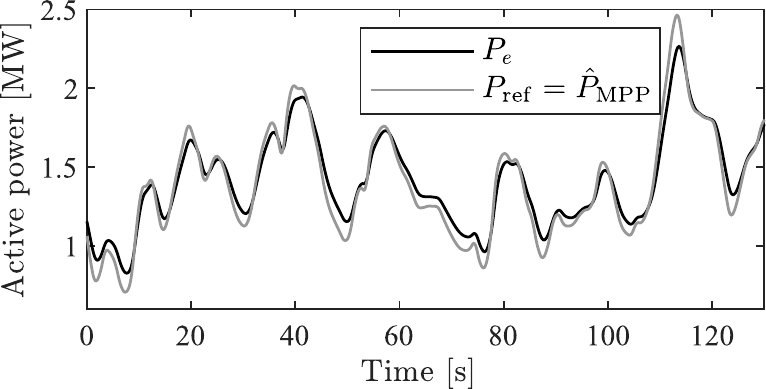}}}
    \caption{Power injection of a MPP tracking WT with recorded wind speed.}
    \label{fig:variable_wind_and_track}
\end{figure}

\begin{figure}[tb!]
    \captionsetup[subfloat]{farskip=0pt}
    \centering
    \subfloat[\label{fig:variable_absolute_step}WT response with and without a~\SI{250}{\kilo\watt} reference step.]
    {{\includegraphics[width=0.49\linewidth]{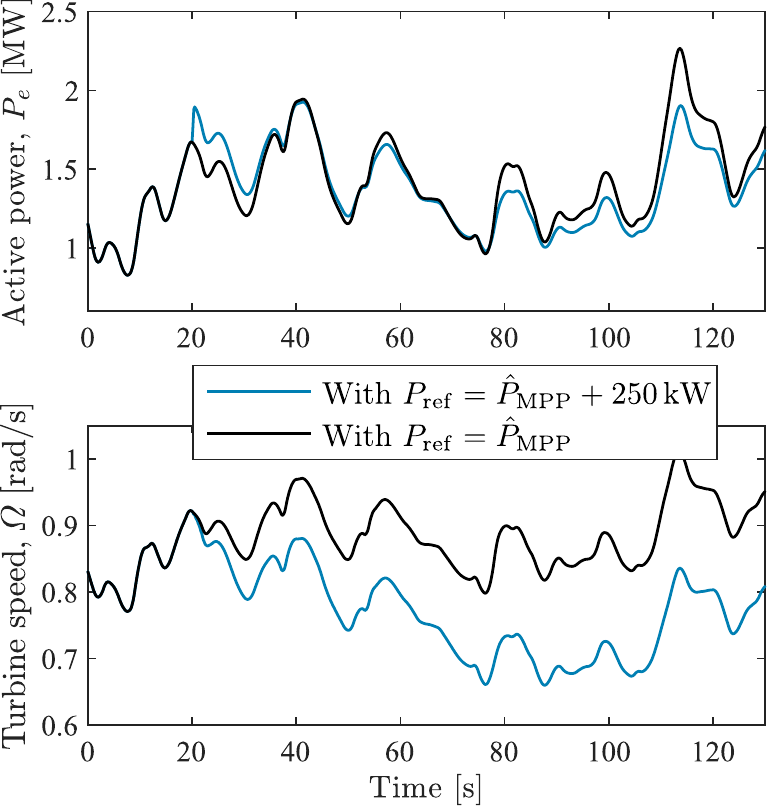}}}
    \hfill
    \subfloat[\label{fig:variable_relative_step}WT response with $\smash{P_\mathrm{ref} = \hat{P}_\mathrm{MPP}}$  as a baseline.]
    {{\includegraphics[width=0.49\linewidth]{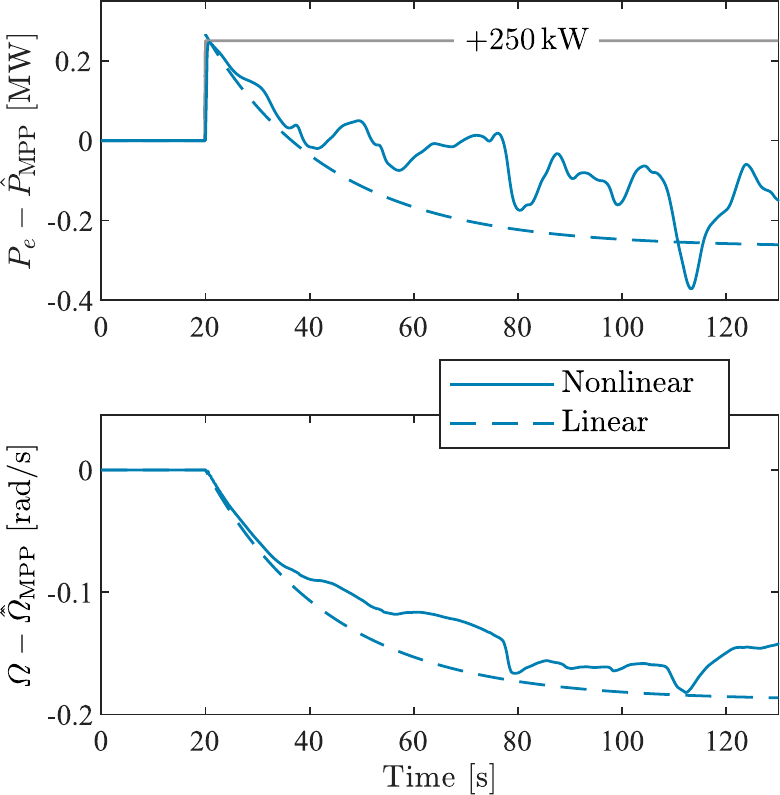}}}
    \caption{Power injection and rotor speed of a MPP tracking WT.}
    \label{fig:variable_step}
\end{figure}

As shown in \cref{fig:variable_wind}, the wind speed may vary significantly in time. The electric power injections of a MPP tracking WT will therefore vary accordingly as shown in \cref{fig:variable_track}. Here we considered a single \SI{5}{\mega\watt} WT operating under recorded wind speed conditions~\cite{syslab}. It is assumed the recorded wind data, $v$, shown in \cref{fig:variable_wind} is uniform across the rotor swept area, and that the measurement
\begin{equation}
    \hat v = \frac{1}{4s +1 }v
\end{equation}
is available to determine the MPP target $\hat{P}_\mathrm{MPP} =  \hat{P}_\mathrm{wind}c_\optimal$. In \cref{fig:variable_absolute_step}, we consider a MPP tracking WT subject to a \SI{250}{\kilo\watt} reference step \SI{20}{\second} into the simulation. We see that the initial power increase is quickly drowned by the output variations resulting from the varying wind speed. However, using the MPP tracking WT with $P_\mathrm{ref} = \hat{P}_\mathrm{MPP}$ as a baseline in \cref{fig:variable_relative_step}, we see that the deterministic WT response to the reference step matches the result at constant wind speed in \cref{fig:openloop_wind_speeds}. The turbine quickly increases its power output, this reduces the rotor speed (compared to the MPP tracking WT) and thus reduces the sustainable output power. 

For the FFR, we will want to control hundreds of WTs at once. The wind speed will vary slightly from one place to another, even for turbines in the same geographical area. Therefore, a realistic simulation study would require stochastic wind data for each individual WT participating in the FFR. This is not practical when we scale up the study, nor does it add any valuable extra insight to the considered control problem since this variability will be present regardless if the WT are used for FFR or not.
As seen in \cref{fig:variable_relative_step}, the deterministic WT response matches the behavior at constant wind speed for a single WT, with some noise due to the varying wind speed. We can assume that the same holds for multiple WTs.
Therefore, we henceforth consider WTs operating at constant wind speeds.

\section{Evaluation in a DVPP and the N5 Test System} \label{sec:PowerSystem}
In this section, we show how the simplified WT model \eqref{eq:linear_plant_approx} can be used for designing FFR. In particular, we will show how the NMP characteristics captured by \eqref{eq:linear_plant_approx} are useful for coordinating FFR from wind with slow FCR from hydro, so that a common FCR and FFR design goal is achieved. 

We do this in two steps. First, we will show how to coordinate wind--hydro resources in a open-loop FCR and FFR setting, creating a DVPP that fulfills design objectives that cannot be fulfilled with wind or hydro individually. 
Finally, we show the implementation of closed-loop FCR and FFR control using feedback from local frequency measurement. We show how FFR from wind resources can be used to fulfill the FCR-D requirements in the hydro-dominated N5 test system.

\subsection{Open-Loop FCR and FFR in a Wind--Hydro DVPP}
\label{sec:DVPP}
\begin{figure}[t!]
		\centering
		\includegraphics[width=\linewidth]{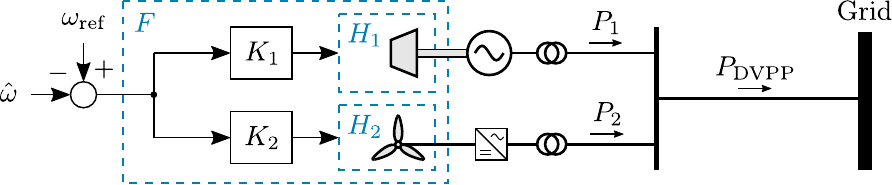}
		\caption{
		 One-line diagram of a wind--hydro DVPP.
		}
		\label{fig:DVPP_block}
\end{figure}

Consider a region with wind and hydro turbines exporting power to the grid as shown in \cref{fig:DVPP_block}. For simplicity, assume that all hydro units are lumped into a single \SI{100}{\mega\voltampere} unit, that the \SI{20}{\mega\watt} wind power park is lumped into a single (scaled) NREL WT, and that the two lumped models are connected to the same bus.

Assuming an initial gate opening $g_0 = 0.8$, servo time constant $T_y = \text{\SI{0.2}{\second}}$, and water time constant $T_\mathrm{w} = \text{\SI{2}{\second}}$, then the linearized hydro model
\begin{equation}
    H_1(s) = 100 \frac{1}{s 0.2 + 1}2 \frac{-s + 0.625}{s+1.25}.
\end{equation}
Let $v=\text{\SI{8}{\meter\per\second}}$, then the linearized WT model
\begin{equation}
    H_2(s) = 4\times 1.8  \frac{s - 0.048}{s+0.048}.
\end{equation}

\subsubsection*{Goal}
Design controllers $K_1$ and $K_2$ so that the hydro unit and wind park behaves like an ideal DVPP fulfilling the FCR requirement,
\begin{equation}
\label{eq:Fideal}
    H_1K_1 + H_2K_2 = F:= 20 \frac{6.5s+1}{(2s+1)(17s+1)}.
\end{equation}

\subsubsection*{Solution} Using the model matching procedure described in~\cite{bjorkDynamicVirtualPowerunpublished,bjorkFundamentalControlPerformance2021}. Design dynamic participation factors $c_1(s)+ c_2(s) = 1$ so that $H_1K_1 + H_2K_2 = F$ is fulfilled for all $s\in\complex$ using the controllers 
\begin{equation}
\label{eq:DVPP_controllers}
    K_i(s) = c_i(s) F(s) /H_i(s),\quad i = 1,\,2.
\end{equation}

For internal stability,  cancellation of NMP zeros is not allowed. Thus, NMP zeros need to be present in $c_1$ and $c_2$.

\begin{enumerate}
    \item 
    % Since the hydro unit will provide the steady state input we start with $c_1$. 
    Starting with the hydro unit, including the NMP zero, let
    \begin{equation}
        c'_1 =\frac{-s + 0.625}{s+0.625}
    \end{equation}
    where we have included a stable pole to make $c'_1$ proper. The location of the pole is a design parameter, but a good starting point is to make $c'_1$ all-pass. 
    % If necessary, add additional poles to make the controllers \eqref{eq:DVPP_controllers} proper.
    
    \item With $c'_1$ defined, the naive approach is to let $c_2' = 1-c_1'$. But since $H_2$ is NMP, this is not possible. Instead, let
    \begin{equation}
    \label{eq:c_step2}
        c_2' = \big(1-c_1' \big)  \frac{s - 0.048}{s+0.048}.
    \end{equation}
    
    \item The sum
    \begin{equation}
        c_1'+c_2' = \frac{s^2 + 0.48s + 0.03}{s^2 + 0.67s + 0.03} \approx 1 + j0
    \end{equation}
    so we could stop here. However, since $c_1' + c_2'$ is stable and MP, perfect matching with $c_1(s)+c_2(s) = 1$ is achievable using 
    \begin{equation}
    \label{eq:DVPP_factors}
        c_1 = \frac{c_1'}{c_1'+c_2'}, \quad \text{and} \quad c_2 = \frac{c_2'}{c_1'+c_2'}.
    \end{equation}
    Controllers $K_1$ and $K_2$ are then obtained by substituting \eqref{eq:DVPP_factors} into \eqref{eq:DVPP_controllers}.
\end{enumerate}

\subsubsection*{Result}
\begin{figure}[t!]
    \captionsetup[subfloat]{farskip=0pt} % {farskip=0pt,nearskip=0pt}
    \centering
    \subfloat[\label{fig:DVPP_line} Power injections.]
    {{\includegraphics[scale=0.52]{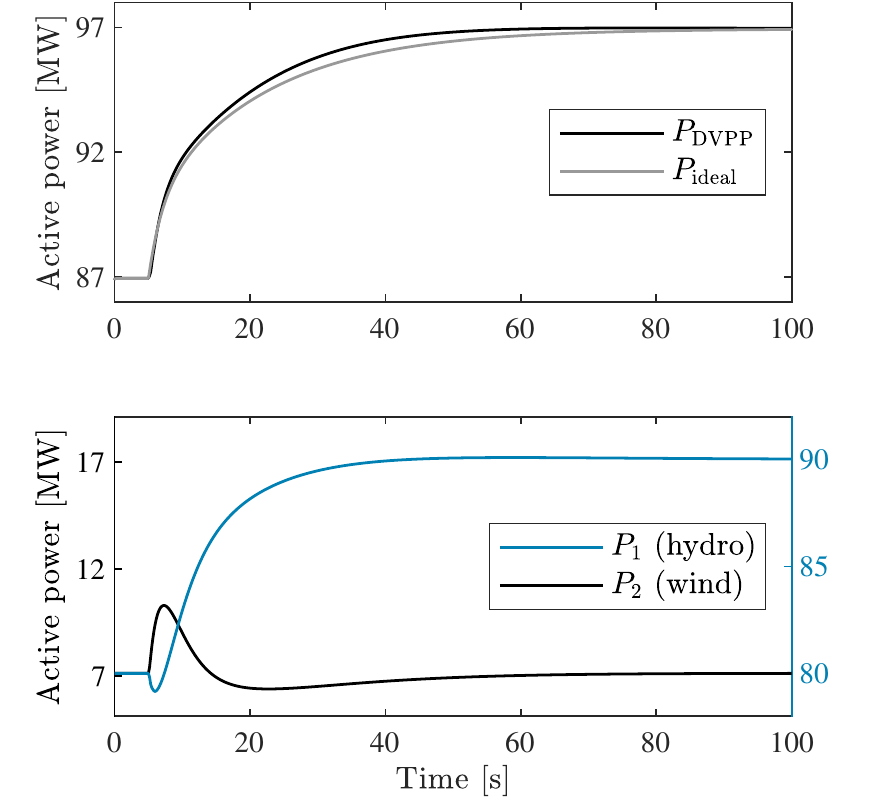}}}
    \hfill
    \subfloat[\label{fig:DVPP_wind} WT response.]
    {{\includegraphics[scale=0.52]{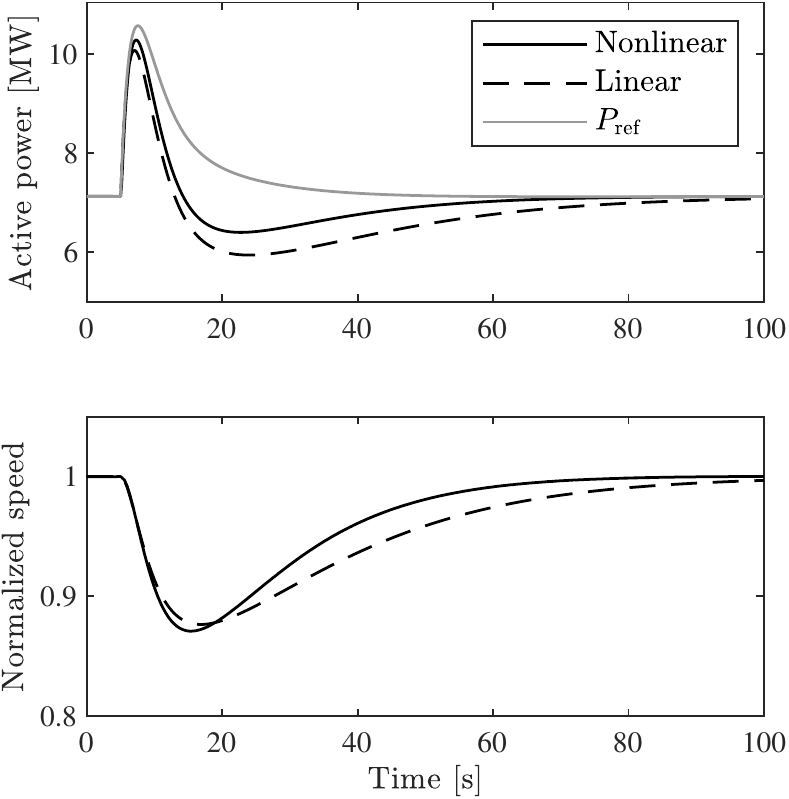}}}
    \caption{Power injection from the wind--hydro DVPP. }
    \label{fig:DVPP}
\end{figure}

The coordinated response to a \SI{0.5}{\hertz} reference step is shown in \cref{fig:DVPP}. As shown in \cref{fig:DVPP_line} the DVPP closely matches the ideal output $P_\mathrm{ideal} = F(s)(\omega_\mathrm{ref}-\hat \omega)$. The discrepancy between the actual response and the ideal response comes from the underestimation of wind power output seen in \cref{fig:DVPP_wind}.

In \eqref{eq:c_step2} we leverage the fact that we can allow the WT to draw power as $|s|\rightarrow 0$. To avoid dynamic interaction, we want the NMP zero of the hydro unit to be much faster than the NMP zero of the WT. Note that the WT zero varies significantly with the turbine speed, but that it is bounded from above \eqref{eq:zero_location}. Thus, the control design, using the derived linearized WT model \eqref{eq:linear_plant_approx}, assumes the worst case interaction.

\subsection{Closed-Loop FCR and FFR in the N5 Test System}
\label{sec:FCR/FFR_in_N5}
Consider the N5 test system introduced in \cref{sec:N5_windhydro}. Assume that the hydro-FCR is complemented with FFR from wind power at buses 2 and 4 as shown in \cref{fig:Nordic_5}.  Assume that the WTs participating in FFR have a total nominal power rating of \SI{2000}{\mega\watt} distributed according to \cref{tab:Nordic5_wind}.
\begin{table}[t!]
\centering
\caption{WT parameters for the \SI{110}{\giga\wattsecond} test case. }
\label{tab:Nordic5_wind}
\begin{tabular}{ccc|cc}
\hline
  Bus & $P_\rated$ [\si{\mega\watt}] & $v$  [\si{\meter\per\second}]& $P_\MPP$ [\si{\mega\watt}] & FFR [\si{\percent}] \\
 \hline
2 & \SI{500} & 10 & 348 & 33 
\\
4 & \SI{1500} & 8 & 534 & 67 
\\
\hline
\end{tabular}
\end{table}

\subsubsection*{Goal} Design coordinated FCR and FFR controllers fulfilling the FCR-D requirements in \cref{sec:theFCR_control_problem}, with feedback from local frequency measurements. That is, design controllers $K_i$, $i=1,\ldots,5$ so that \eqref{eq:asmFCRD} is fulfilled, with
\begin{equation}
    \sum_{i=1}^5 H_i(s) K_i(s)  =  R_\mathrm{FCR} \frac{6.5s+1}{(2s+1)(17s+1)}.
\end{equation}

\subsubsection*{Solution} Using the three-step model matching approach in \cref{sec:DVPP},
dynamic participation factors $c_i(s)$ are designed so that $\sum_{i=1}^5c_i(s) = 1$, $\forall s\in \complex$. Let the shares of FCR and FFR be distributed according to \cref{tab:Nordic5} and \cref{tab:Nordic5_wind}, respectively. The resulting controllers $K_i(s) = c_i(s) F_\mathrm{FCR}(s) /H_i(s)$ will be fourth order. For more details and analysis of the design procedure see~\cite{bjorkDynamicVirtualPowerunpublished,bjorkFundamentalControlPerformance2021}.

\subsubsection*{Result}
\begin{figure}[t!]
    \captionsetup[subfloat]{farskip=0pt}
    \centering
    \subfloat[\label{fig:Nordic5_wind_and_hydro} FCR and FFR response.]
    {{\includegraphics[scale=0.52]{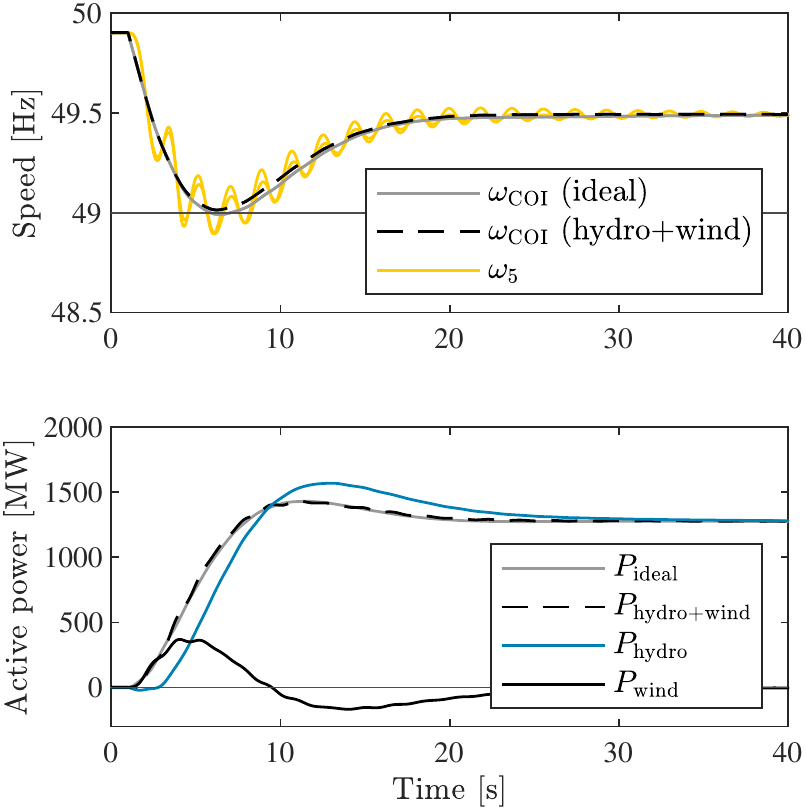}}}
    \hfill
    \subfloat[\label{fig:Nordic5_} WT output and speed.]
    {\includegraphics[scale=0.52]{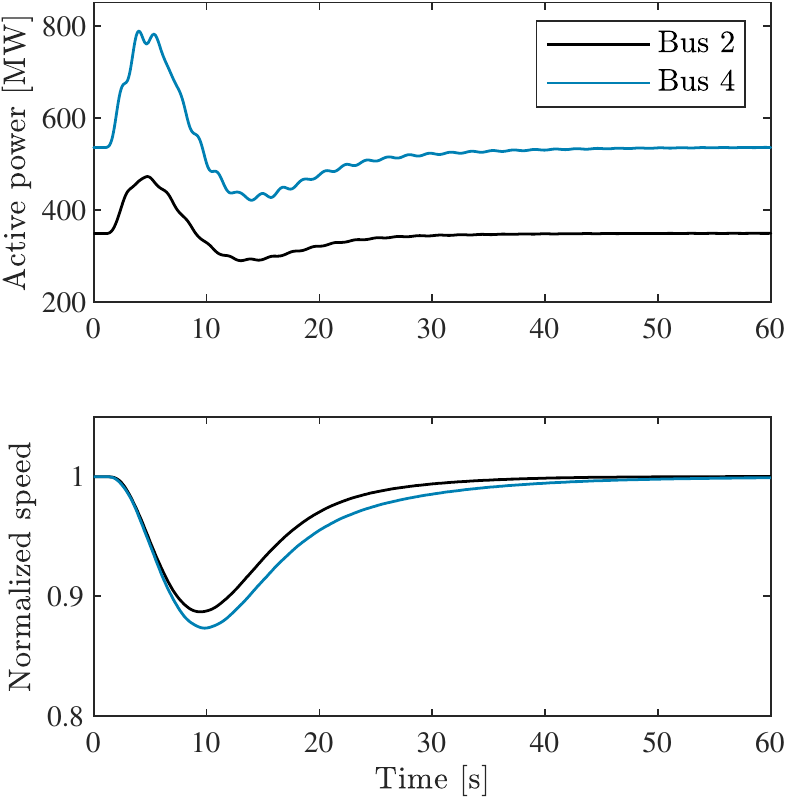}}
    \caption{Response to a \SI{1400}{\mega\watt} fault with coordinated FCR and FFR.}
    \label{fig:Nordic5_2}
\end{figure}

The system response following the dimensioning fault, i.e., the disconnection of the \SI{1400}{\mega\watt} NordLink dc cable, is shown in \cref{fig:Nordic5_2}. Comparing \cref{fig:Nordic5_hydro_1} with \cref{fig:Nordic5_wind_and_hydro} we see that the transient frequency response is improved when the hydro-FCR, $P_\mathrm{hydro}$, is supported by, $P_\mathrm{wind}$, from WTs at buses 2 and 4. The combination $P_\mathrm{hydro+wind}= P_\mathrm{hydro}+P_\mathrm{wind}$ matches the ideal response $P_\mathrm{ideal} = F_\mathrm{FCR}(s)(\omega_\mathrm{ref}-\hat \omega)$. With the nadir kept above \SI{49}{\hertz}, the FCR-D requirement are now fulfilled. By utilizing the WTs for FFR, this is achieved without increasing the bandwidth of the hydro units. Thus, the frequency response is improved without reducing the closed-loop stability margin~\cite{agneholmFCRDDesignRequirements2019}.

In \cref{fig:Nordic5_} we see that the WTs participating in FFR decelerate to around \SI{90}{\percent} of optimal speed following the fault. Since the WT based on the NREL model allows for a deceleration to around \SI{80}{\percent} of optimal speed, there is more room for FFR improvement. Note however that the \SI{500}{\mega\watt} WT at bus 2 almost reaches the rated power. If the limit is reached, the WT will saturate. However, the WT at bus 4 has plenty of margin left. This shows how only a limited wind resource is enough to compensate for the FCR bandwidth limitation of hydro power in this power system. In fact, the FCR-D requirements can be fulfilled with significantly less wind resources.

\subsubsection*{Sensitivity Study with \SI{50}{\percent} Wind Resources}
\begin{figure}[t!]
    \captionsetup[subfloat]{farskip=0pt}
    \centering
    \subfloat[\label{fig:Nordic5_wind_and_hydro_x2} FCR and FFR response.]
    {{\includegraphics[scale=0.52]{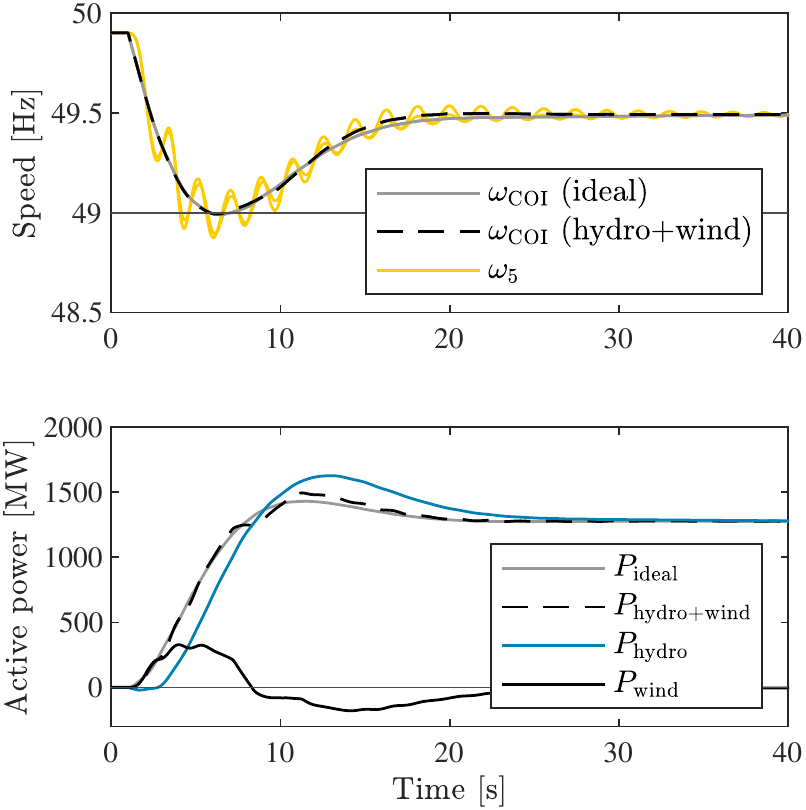}}}
    \hfill
    \subfloat[\label{fig:Nordic5_x2} WT output and speed.]
    {\includegraphics[scale=0.52]{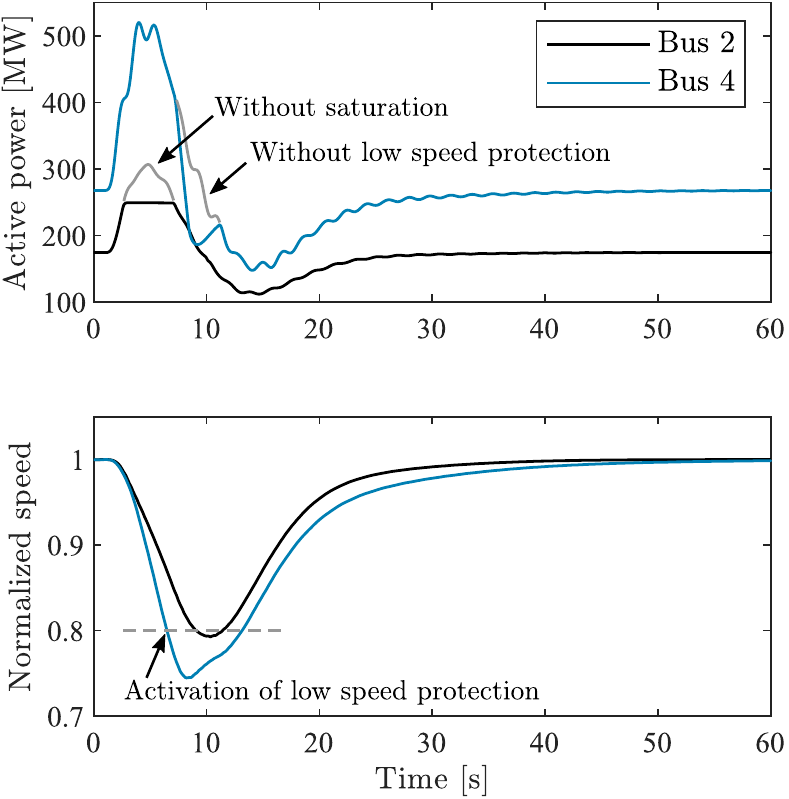}}
    \caption{Sensitivity study with \SI{50}{\percent} of the wind resources in \cref{fig:Nordic5_2}.}
    \label{fig:Nordic5_2_x2}
\end{figure}

Consider the same scenario shown in \cref{fig:Nordic5_2} but now  let $P_\mathrm{nom} = \text{\SI{250}{\mega\watt}}$ at bus 2 and $P_\mathrm{nom} = \text{\SI{750}{\mega\watt}}$ at bus 4. As shown in \cref{fig:Nordic5_x2}, the WT at bus 2 now reaches its saturation limit. This has a small effect on the nadir in \cref{fig:Nordic5_wind_and_hydro_x2}, but not enough to violate the FCR-D requirement. With less wind resources, the FFR response gives a larger deceleration of the participating WTs. {As seen in \cref{fig:Nordic5_x2}, the low speed protection \cref{eq:low-speed-prot} is activated at bus 4. This disables the FFR input and switches to a stabilizing controller with higher gain. The low speed protection is active until the measured speed at bus 4 is restored to above 0.8 or (as is the case in \cref{fig:Nordic5_x2}) until the desired set point falls below the low speed protection set point.}

\section{Conclusions} \label{sec:Conclusion}
A variable-speed controlled WT model based on the NREL \SI{5}{\mega\watt} baseline WT model has been presented. The new model-based variable-speed feedback controller allows the WT to provide temporary FFR when operated at the MPP, without the need for curtailment. The controller has been designed to give the WT similar dynamic properties for various wind speeds. This facilitated a meaningful linearization of the WT, resulting in a first-order linearized model, capturing the relevant dynamic properties for FFR control design at various wind speeds. {Having a suitable linearization allows us to use tools from linear control theory to determine if the frequency control objective is feasible with the available actuators. The simplified linear WT model is useful for designing controllers to coordinate wind-FFR with slower FCR. As a challenging case study, we considered the coordination of wind-FFR with hydro-FCR. The coordinated FCR and FFR can fulfill dynamic design objectives not achievable by wind or hydro individually.} This was shown in a local DVPP and in a 5-machine representation of the Nordic synchronous grid. During FFR, WTs take the lead (via synthetic inertia) in order to limit the frequency nadir and compensate for the initial under production of hydro units, as depicted in \cref{fig:Nordic5_hydro_zoom}. After this initial transitory, hydro units take over during FCR recovery, compensating not only for the initial down frequency event, but also for the kickback effect caused by the WT-based FFR. The kickback effect is characterized by the NMP zero in \cref{eq:closed_loop_exact}. Including this NMP characteristic in the coordinated FCR and FFR control design, results in an excellent frequency recovery as depicted in \cref{fig:Nordic5_wind_and_hydro,fig:Nordic5_wind_and_hydro_x2}, compared to hydro-only FCR in \cref{fig:Nordic5_hydro_1}.

Wind power is inherently stochastic. This holds true regardless if the WT is participating in FFR or not.  In \cref{sec:vary_speed} we showed that the deterministic response of wind-FFR at variable wind speeds matches the behavior at constant wind speed. A power outtake above the MPP decelerates the rotor and therefore reduces the sustainable power output. To ensure stable operation, the WT is equipped with a low speed protection that bypasses the FFR control. The protection may be triggered if the requested energy outtake is too large. The best way to avoid this is to design the FCR-D design target with some margin, and to have more turbines participating in the FFR. 
It is shown that the dynamic frequency requirements can be achieved even if some of the WTs participating in the FFR saturates or enter low speed protection mode.

The intermittency of wind power is an issue for the SO both for the steady power production and when guaranteeing the reserve power available for FFR. As argued in \cref{sec:vary_speed} this problem is partly solved by controlling a larger fleet of WTs, since local variations in wind speed will become less relevant. Future work will study how to account for this intermittency when committing WTs to FFR support. We will also compare the reliability of FFR with the MPP tracking control scheme to a curtailed control scheme.

In this work, only the NREL \SI{5}{\mega\watt} WT model have been considered. But as indicated in \cref{rem:scalability}, the dynamic properties should be fairly consistent (in \si{\perunit}) for WTs of various types and sizes. Future work will extend the analysis to other WT types.

\appendix
% \subsection{Choosing the FCR-D Design Target}
\label{appendix:tuning_target}
The first guess for the FCR-D design target \eqref{eq:asmFCRD} is motivated by the SO's requirements.
Following an immediate frequency fall from 49.9 to \SI{49.5}{\hertz}, the reserves should be \SI{50}{\percent} activated within \SI{5}{\second}~\cite{entso-eNordicSynchronousArea2018}. We therefore let  
\begin{equation}
    \label{eq:FCR_temp}
        F_\mathrm{temp}(s) = R_\mathrm{FCR} \frac{1}{sT_\mathrm{temp}+1}
    \end{equation} 
    with the time constant 
    \begin{equation}
    \label{eq:T_temp}
        T_\mathrm{temp} = -5/\ln(0.50) \approx 7.2.
    \end{equation}
Considering a \SI{1400}{\mega\watt} dimensioning fault and \SI{400}{\mega\watt/\hertz} inherent load damping we let the gain 
    \begin{equation}
        R_\mathrm{FCR} = {1400}/({49.9-49.5}) - 400 = 3100,
    \end{equation}
so that the frequency stabilizes at \SI{49.5}{\hertz}.
    At this point, the disturbance response
    \begin{equation}
    \label{eq:dist_response}
        \frac{1}{s 2 W_\mathrm{kin} /50 + 400 + F_\mathrm{temp}(s)}
    \end{equation}
    fulfills the FCR-D requirements. With $W_\mathrm{kin} = \text{\SI{110}{\giga\wattsecond}}$, the nadir reaches almost \SI{49.0}{\hertz} following a \SI{1400}{\mega\watt} load step. This means that \cref{eq:T_temp} is the slowest possible time constant that we may allow with the first-order design target \cref{eq:FCR_temp}. One problem is that even if the target \cref{eq:FCR_temp} can be perfectly realized, the system will overshoot and oscillate when restoring the system frequency. This can be amended by adding a lead-filter to \cref{eq:FCR_temp} and readjusting the time constant \cref{eq:T_temp} so that the disturbance response \cref{eq:dist_response} still reaches a nadir of \SI{49.0}{\hertz}. This does not significantly change the FCR activation time, but gives a more desirable dynamic response. For the N5 test system in \cref{sec:N5_windhydro}, the second-order transfer function \cref{eq:asmFCRD} is a suitable design target, as shown in \cref{fig:Nordic5_hydro_1}.

%\pagebreak
%\listoffigures

%\appendix

%%%%%%%%%%%%%%%%%%%%%%%%%%%%%%%%%%%%%%%%%%%%%%%%%%%%%%%%%%%%

\bibliographystyle{IEEEtran}

% \bibliography{myRefs} %\bibliography{C:/Users/joakbj/Zotero/myRefs}
% LETS START A NEW MANUAL .bib file
% \bibliography{manual_bib.bib}
\bibliography{bib_IEEE}

\begin{IEEEbiography}[{\includegraphics[width=1in,height=1.25in,clip,keepaspectratio]{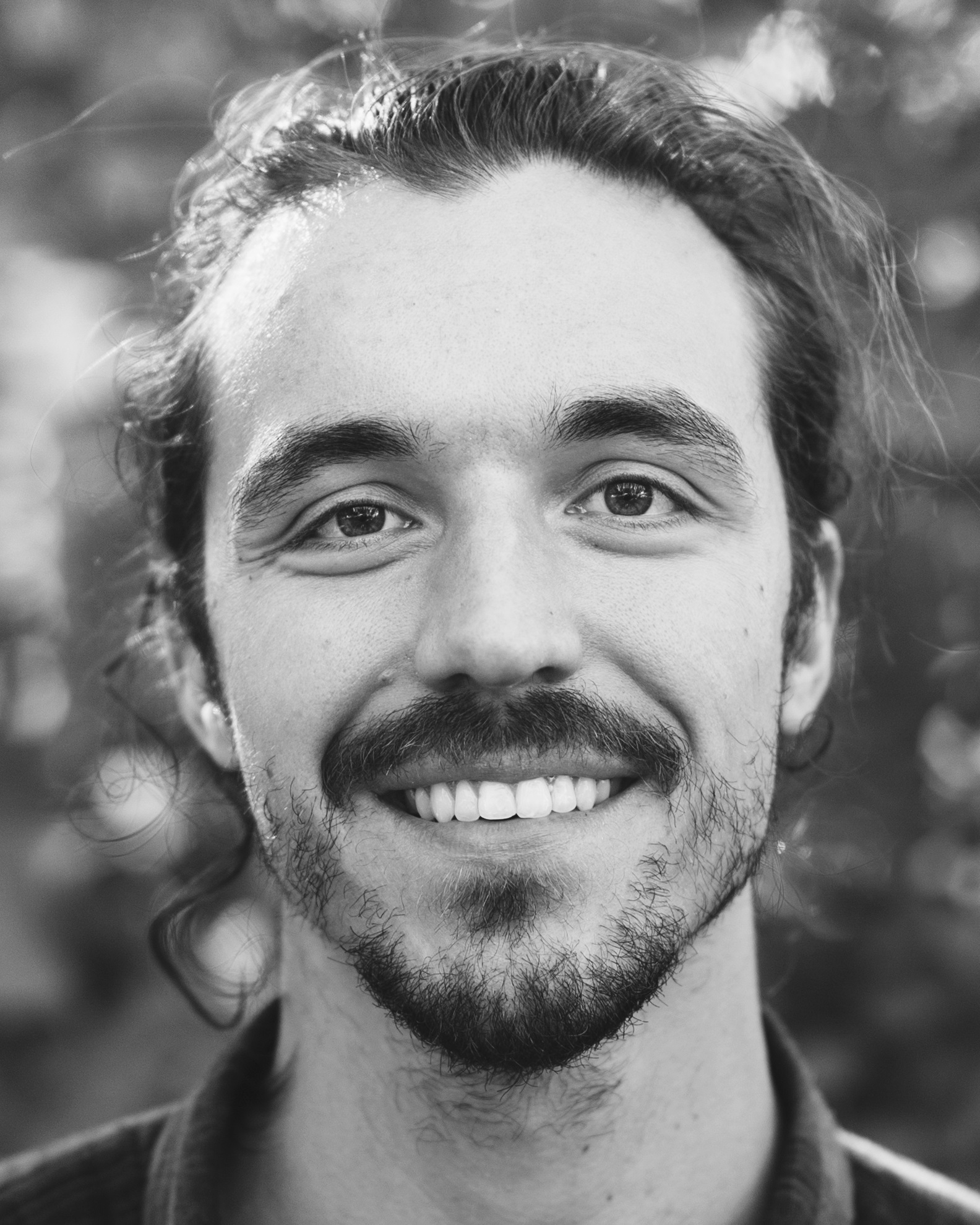}}]{Joakim Bj\"ork} 
	(S'17--M'21)
	received the M.Sc.\ degree in energy systems engineering from Uppsala University, Uppsala, Sweden, in collaboration with Swedish University of Agricultural Sciences, Uppsala, Sweden, in 2016.
	He obtained the Ph.D.\ degree in electrical engineering at the Division of Decision and Control Systems, KTH Royal Institute of Technology, Stockholm, Sweden, in 2021.
	
	In 2021, he joined the Swedish National Grid,
	Department of Power Systems. His research interests are fundamental control limitations in networked control systems with applications in frequency and voltage control of ac-dc power systems.
\end{IEEEbiography}

\begin{IEEEbiography}[{\includegraphics[width=1in,height=1.25in,clip,keepaspectratio]{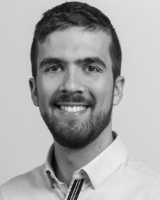}}]{Daniel~V\'azquez~Pombo}
	(S'19) Received the M.Sc.\ degree in Electric Power Systems from Aalborg University, Denmark, in 2018. Since 2019, he is part of Vattenfall R\&D where he is the head of Control and Optimization; a group specialised in the development of advanced applications for Utility-Scale Hybrid Power Plants.
	
	He is currently pursuing a Ph.D.\ in Energy Management of Isolated Hybrid Power Systems at the Technical University of Denmark. His research interests include power system optimization, operation and control, frequency stability, renewable energy integration, and data analytics.	
\end{IEEEbiography}

\vfill \newpage
\begin{IEEEbiography}[{\includegraphics[width=1in,height=1.25in,clip,keepaspectratio]{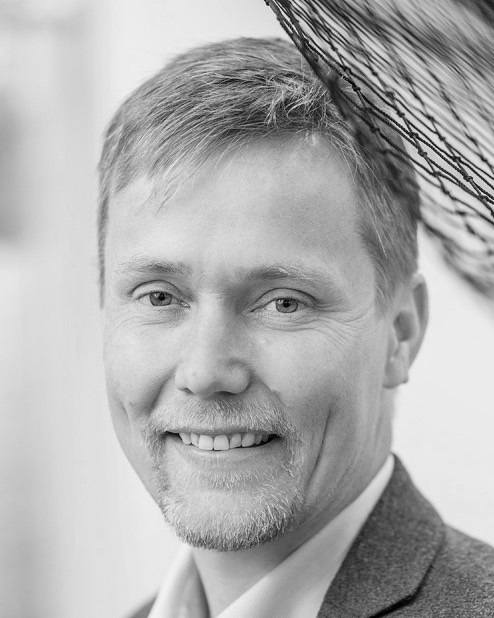}}]{Karl Henrik Johansson} 
	(F'13) 
	Karl H. Johansson is Professor with the School of Electrical Engineering and Computer Science at KTH Royal Institute of Technology in Sweden and Director of Digital Futures. He received M.Sc.\ and Ph.D.\ degrees from Lund University. 
	
	He has held visiting positions at UC Berkeley, Caltech, NTU, HKUST Institute of Advanced Studies, and NTNU. His research interests are in networked control systems and cyber-physical systems with applications in transportation, energy, and automation networks. He is a member of the Swedish Research Council's Scientific Council for Natural Sciences and Engineering Sciences. He has served on the IEEE Control Systems Society Board of Governors, the IFAC Executive Board, and is currently Vice-President of the European Control Association. He has received several best paper awards and other distinctions from IEEE, IFAC, and ACM. He has been awarded Distinguished Professor with the Swedish Research Council and Wallenberg Scholar with the Knut and Alice Wallenberg Foundation. He has received the Future Research Leader Award from the Swedish Foundation for Strategic Research and the triennial Young Author Prize from IFAC. He is Fellow of the IEEE and the Royal Swedish Academy of Engineering Sciences, and he is IEEE Control Systems Society Distinguished Lecturer.
\end{IEEEbiography}

\vfill

\end{document}